\documentclass[aps,floatfix,superscriptaddress,twocolumn]{revtex4-1}

\usepackage[dvips]{graphicx}

\usepackage{amssymb,amsfonts,amsmath}
\usepackage{hyperref}
\usepackage{cancel}
\usepackage{natbib}

\begin{document}

\title{Thermodynamic glass transition in a spin glass without time-reversal symmetry}

\author{R.~A.~Ba\~nos} \affiliation{Instituto de Biocomputaci\'on y
  F\'{\i}sica de Sistemas Complejos (BIFI), 50009 Zaragoza, Spain.}
  \affiliation{Departamento
  de F\'\i{}sica Te\'orica, Universidad
  de Zaragoza, 50009 Zaragoza, Spain.} 

\author{A.~Cruz} \affiliation{Departamento
  de F\'\i{}sica Te\'orica, Universidad
  de Zaragoza, 50009 Zaragoza, Spain.} \affiliation{Instituto de Biocomputaci\'on y
  F\'{\i}sica de Sistemas Complejos (BIFI), 50009 Zaragoza, Spain.}

\author{L.A.~Fernandez} \affiliation{Departamento
  de F\'\i{}sica Te\'orica I, Universidad
  Complutense, 28040 Madrid, Spain.} \affiliation{Instituto de Biocomputaci\'on y
  F\'{\i}sica de Sistemas Complejos (BIFI), 50009 Zaragoza, Spain.}

\author{J.~M.~Gil-Narvion} \affiliation{Instituto de Biocomputaci\'on y
  F\'{\i}sica de Sistemas Complejos (BIFI), 50009 Zaragoza, Spain.}

\author{A.~Gordillo-Guerrero}\affiliation{D. de  Ingenier\'{\i}a
El\'ectrica, Electr\'onica y Autom\'atica, U. de Extremadura,
  10071, C\'aceres, Spain.}\affiliation{Instituto de Biocomputaci\'on y
  F\'{\i}sica de Sistemas Complejos (BIFI), 50009 Zaragoza, Spain.}

\author{M.~Guidetti}  \affiliation{Instituto de Biocomputaci\'on y
  F\'{\i}sica de Sistemas Complejos (BIFI), 50009 Zaragoza, Spain.}

\author{D.~I\~niguez} \affiliation{Instituto de Biocomputaci\'on y
  F\'{\i}sica de Sistemas Complejos (BIFI), 50009 Zaragoza, Spain.}
  \affiliation{Fundaci\'on ARAID, Diputaci\'on General de Arag\'on,
  Zaragoza, Spain}

\author{A.~Maiorano} 
  \affiliation{Dipartimento di Fisica, Universit\`a di Roma ``La Sapienza'', 00185 Roma, Italy.}\affiliation{Instituto de Biocomputaci\'on y
  F\'{\i}sica de Sistemas Complejos (BIFI), 50009 Zaragoza, Spain.}

\author{E.~Marinari}\affiliation{Dipartimento di Fisica, Universit\`a di Roma ``La Sapienza'', IPCF-CNR and INFN, 00185 Roma, Italy.}

\author{V.~Martin-Mayor} \affiliation{Departamento de F\'\i{}sica
  Te\'orica I, Universidad Complutense, 28040 Madrid, Spain.} \affiliation{Instituto de Biocomputaci\'on y
  F\'{\i}sica de Sistemas Complejos (BIFI), 50009 Zaragoza, Spain.}

\author{J.~Monforte-Garcia} \affiliation{Instituto de Biocomputaci\'on y
  F\'{\i}sica de Sistemas Complejos (BIFI), 50009 Zaragoza, Spain.}
  \affiliation{Departamento
  de F\'\i{}sica Te\'orica, Universidad
  de Zaragoza, 50009 Zaragoza, Spain.}

\author{A.~Mu\~noz Sudupe} \affiliation{Departamento
  de F\'\i{}sica Te\'orica I, Universidad
  Complutense, 28040 Madrid, Spain.} 

\author{D.~Navarro} \affiliation{D.  de Ingenier\'{\i}a,
  Electr\'onica y Comunicaciones and I3A, U. de Zaragoza, 50018 Zaragoza, Spain.}

\author{G.~Parisi}\affiliation{Dipartimento di Fisica, Universit\`a di Roma ``La Sapienza'', IPCF-CNR and INFN, 00185 Roma, Italy.}

\author{S.~Perez-Gaviro} \affiliation{Instituto de Biocomputaci\'on y
  F\'{\i}sica de Sistemas Complejos (BIFI), 50009 Zaragoza, Spain.}

\author{J.~J.~Ruiz-Lorenzo} \affiliation{Departamento de
  F\'{\i}sica, Universidad de Extremadura, 06071 Badajoz, Spain.}\affiliation{Instituto de Biocomputaci\'on y
  F\'{\i}sica de Sistemas Complejos (BIFI), 50009 Zaragoza, Spain.}

\author{S.F.~Schifano} \affiliation{Dipartimento
  di Fisica, Universit\`a di Ferrara and INFN - Sezione di Ferrara,
  Ferrara, Italy.} 

\author{B.~Seoane} \affiliation{Departamento de F\'\i{}sica
  Te\'orica I, Universidad Complutense, 28040 Madrid, Spain.} \affiliation{Instituto de Biocomputaci\'on y
  F\'{\i}sica de Sistemas Complejos (BIFI), 50009 Zaragoza, Spain.}

\author{A.~Tarancon} \affiliation{Departamento
  de F\'\i{}sica Te\'orica, Universidad
  de Zaragoza, 50009 Zaragoza, Spain.} \affiliation{Instituto de Biocomputaci\'on y
  F\'{\i}sica de Sistemas Complejos (BIFI), 50009 Zaragoza, Spain.}

\author{P.~Tellez} \affiliation{Departamento
  de F\'\i{}sica Te\'orica, Universidad
  de Zaragoza, 50009 Zaragoza, Spain.} 

\author{R.~Tripiccione} \affiliation{Dipartimento
  di Fisica, Universit\`a di Ferrara and INFN - Sezione di Ferrara,
  Ferrara, Italy.} 
 
\author{D.~Yllanes}  \affiliation{Departamento de F\'\i{}sica
  Te\'orica I, Universidad Complutense, 28040 Madrid, Spain.}\affiliation{Instituto de Biocomputaci\'on y
  F\'{\i}sica de Sistemas Complejos (BIFI), 50009 Zaragoza, Spain.}

\date{\today}

\begin{abstract} 
Spin glasses are a longstanding model for the sluggish dynamics that appears
at the glass transition. However, spin glasses differ from structural glasses
for a crucial feature: they enjoy a time reversal symmetry. This symmetry can
be broken by applying an external magnetic field, but embarrassingly little is
known about the critical behaviour of a spin glass in a field. In this
context, the space dimension is crucial. Simulations are easier to interpret
in a large number of dimensions, but one must work below the upper critical
dimension (i.e., in $d<6$) in order for results to have relevance for
experiments. Here we show conclusive evidence for the presence of a phase
transition in a four-dimensional spin glass in a field. Two ingredients were
crucial for this achievement: massive numerical simulations were carried out
on the Janus special-purpose computer, and a new and powerful finite-size
scaling method.
\end{abstract}
\maketitle

The glass transition differs from standard phase transitions in that the
equilibration time of glass formers (supercooled liquids, polymers, proteins,
superconductors, etc.) diverges without dramatic changes in their structural
properties \cite{debenedetti:97,debenedetti:01,cavagna:09}.  The
reconciliation of the dynamic slowdown with the apparent immutability of
glass formers is a major challenge for condensed matter physics.

Spin glasses (which are disordered magnetic alloys \cite{mydosh:93}) enjoy a
privileged status in this context, as they provide the simplest model system
both for theoretical and experimental studies of a glassy dynamics. On the
experimental side, time-dependent magnetic fields provide a wonderful tool to
probe the dynamic response, which can be accurately measured with a SQUID (for 
instance, see \cite{herisson:02}). On
the theoretical side, magnetic systems are notably easier to model and to
simulate numerically. In fact, special-purpose computers have been built for
the simulation of spin glasses \cite{cruz:01,ogielski:85,janus:08, janus:08b}.

Yet, spin glasses differ from most glassy systems in a crucial
feature: like all magnetic systems, they enjoy time-reversal symmetry
in the absence of an applied magnetic field. In fact, we now know that
their glassy dynamics is due to a bona fide phase transition in which
the time-reversal symmetry is spontaneously
broken \cite{gunnarsson:91,ballesteros:00,palassini:99}.
Yet, in the
presence of an applied magnetic field, the experimental spin-glass
dynamics is just as glassy, although the field explicitly breaks the
symmetry.

However, whether spin glasses in a magnetic field undergo a phase transition
has been a long-debated and still open question (see
\cite{bray:11,parisi:11} for recent, opposed views). In the
mean-field approximation, which is valid for large spatial dimension down to
the upper critical dimension $d_\mathrm{u}=6$ \cite{bray:80}, the de
Almeida-Thouless line~\cite{dealmeida:78} separates the high-temperature paramagnetic phase from
the glassy phase~\footnote{A more careful analysis
is needed in order to reach the same
conclusion in the range $6<d<8$ \cite {fisher:85}.}. 
 Yet, recent numerical
simulations in spatial dimensions below $d_\mathrm{u}$ did not find the
transition in a field \cite{young:04,jorg:08b}. Experimental studies have been
conducted as well, with conflicting
conclusions \cite{jonson:05,petit:99,petit:02,tabata:10}.
In spite of this, it
has been argued that the would-be spin-glass transition in a magnetic field
sets the universality class for the thermodynamic glass
transition \cite{moore:02}.

Here, we present conclusive evidence for a spin-glass transition in
the presence of an external magnetic field in the four-dimensional
Edwards-Anderson model (hence, well below $d_\mathrm{u}$). This result
was obtained by means of a large-scale numerical simulation, partly
carried out on the Janus computer \cite{janus:08}.
Due to some
pathologies of the spin-glass correlation function \cite{leuzzi:09}, our
analysis method departs from the standard one. We compute critical
exponents, widely differing from the zero field case, with an accuracy
of five percent. The failure of previous work to identify the
transition is explained in terms of very strong corrections to
scaling.

\section{Results}
We consider the Edwards-Anderson model with Ising spins ($S_{\boldsymbol x}=\pm1$) sitting on 
the nodes of a $D=4$ cubic lattice of size $V=L^D$. Our  Hamiltonian is
\begin{equation}
\mathcal H = - \sum_{\langle \boldsymbol x, \boldsymbol y\rangle} J_{\boldsymbol x\boldsymbol y} S_{\boldsymbol x}S_{\boldsymbol y}-h \sum_{\boldsymbol x} S_{\boldsymbol x},
\end{equation}
where $\langle \boldsymbol x, \boldsymbol y\rangle$ indicates that the
sum is taken over all nearest-neighbour pairs and each $J_{\boldsymbol
x\boldsymbol y}$ is $\pm1$ with $50\%$ probability. We provide details
about our numerical simulations in Appendix~\ref{sec:simulation-details}.

As stated in the introduction, we want to investigate whether this
system experiences a second-order phase transition in the presence of
a non-zero magnetic field $h$. This is typically checked through the
study of some correlation length $\xi$, which is a good marker of
the scale invariance commonly associated to continuous transitions.

To this end, we begin by defining the spatial autocorrelation function
$G(\boldsymbol r)$.  This can actually be done in several ways in the presence
of a magnetic field (see Appendix~\ref{sec:correlation-functions} for details).
Then, $\xi$ is just the characteristic length for the long-distance decay of
$G(\boldsymbol r)$. In order to arrive at an appropriate definition for finite
lattice systems, one typically considers the propagator in Fourier space,
$\hat G(\boldsymbol k)$, and defines the second-moment correlation length
$\xi_2$ from a truncated Ornstein-Zernike expansion ---eqs.~\eqref{eq:OZ}
and~\eqref{eq:xi2}.
\begin{figure}[t]
\centering
\includegraphics[height=\columnwidth,angle=270]{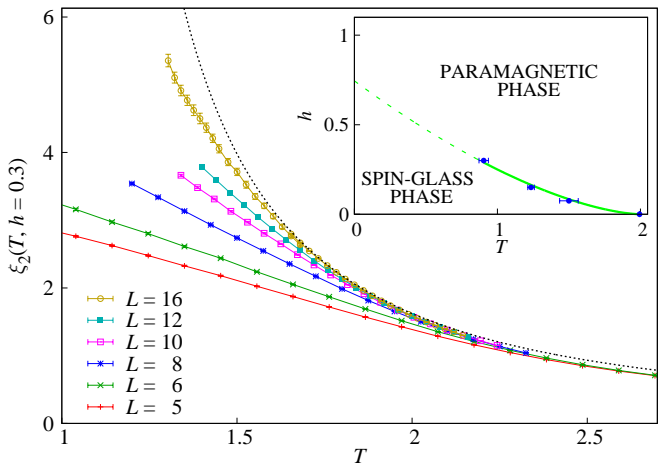}
\caption{Plot of the second moment correlation length $\xi_2$
 ---eq.~\eqref{eq:xi2}---
against temperature in an external field $h=0.3$. 
There is a clear crossover from the  convergence to a  finite envelope
at high $T$ to the more rapid growth at low $T$. As this paper shows,
this is caused by the onset of a spin-glass transition. The dotted black line is a fit to a critical divergence as $\xi_2^\infty\propto[T-T_\text{c}(h)]^{-\nu}$, where $T_\text{c}$ and $\nu$ are taken from 
Table~\ref{tab:parameters}.
The inset is a sketch of the phase diagram (the de Almeida-Thouless line),
including a fit to the Fisher-Sompolinsky scaling
$h^2_\text{c}(T) \simeq A |T-T_\text{c}^{(0)}|^{\beta^{(0)}+\gamma^{(0)}}$~\cite{fisher:85}.
The quantities with a superindex $(0)$ are the values for the $h=0$ critical point,~\cite{jorg:08c,marinari:99b}
so the only free parameter is the amplitude $A$.
\label{fig:xi}}
\end{figure}

We have plotted $\xi_2$ in Figure~\ref{fig:xi} for all our lattice sizes
and $h=0.3$. There is a clear change of regime from the 
high-temperature behaviour, where we can see a finite enveloping curve, 
to the growth of the correlation length at low temperatures. We intend to show
that this change of regime actually corresponds to a phase transition, using
finite-size scaling \cite{amit:05}.
\begin{figure}[t]
\centering
\includegraphics[height=\linewidth,angle=270]{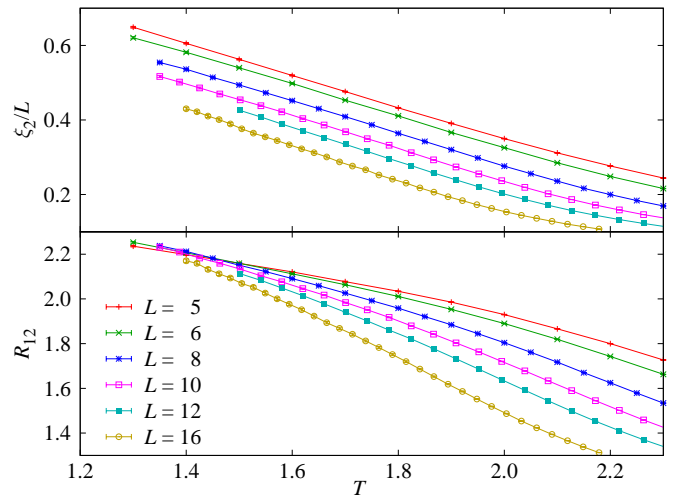}
\caption{Top: plot of $\xi_2/L$ as a function of temperature
for all our lattice sizes at $h=0.15$. According to leading-order 
finite-size scaling, the curves for different sizes should
intersect at the phase transition point, but this behaviour
is not seen in the plot. This apparent lack of scale invariance
 has led some authors to conclude
that there is no phase transition in this system.
Bottom: Same plot of the dimensionless ratio $R_{12}$, eq.~\eqref{eq:R12},
which should have the same leading-order scaling as $\xi_2/L$. Unlike the 
correlation length, however, $R_{12}$ does exhibit very clear intersections,
signalling the presence of a second-order phase transition. The dramatic
improvement in the scaling, compared to the top panel, is explained by
 the pernicious effect on $\xi_2$ of the anomalous behaviour 
in the correlation function for zero momentum.
\label{fig:xi-R12}}
\end{figure} 

In principle, at the transition point there should be scale
invariance in the system, meaning that
\begin{equation}\label{eq:xi-FSS}
\xi_2/L = f_\xi\bigl(L^{1/\nu} t\bigr) + \ldots,\quad t = \frac{T-T_\text{c}(h)}{T_\text{c}(h)}
\end{equation}
where $\nu$ is the thermal critical exponent and the dots represent
corrections to leading scaling, expected to be unimportant for large
lattice sizes.  Therefore, the curves of $\xi_2/L$ for large lattices
should intersect at the critical point $t=0$. Previous attempts to find
$T_\text{c}$ using this approach, however, have generally concluded
that these intersections cannot be found (or, rather, that the
apparent intersection point goes to $T=0$ as $L$
grows) \cite{young:04,jorg:08b}. Indeed, if we look at the top
panel of Figure~\ref{fig:xi-R12}, we see that  either there is no
phase transition or $\xi_2$ is completely in a preasymptotic regime.

Some authors, working with $D=1$ models with long-range interactions,
have already offered an explanation for this apparent lack of scale
invariance: the propagator behaves anomalously, but only for the
$\boldsymbol k=0$ mode \cite{leuzzi:09}. This results in very strong corrections to the
leading scaling term of eq.~\eqref{eq:xi-FSS}, since
the second-moment correlation length depends on $\hat G(\boldsymbol k=0)$. 
We have checked numerically that this phenomenon is also at play in our $D=4$
system, which is probably a general consequence of 
the presence of Goldstone bosons in the system (see
Appendix~\ref{sec:correlation-functions} for a discussion of this
phenomenon). 

In order to avoid this issue, in this paper we take a novel approach,
eschewing $\xi_2/L$ in favour of a new dimensionless ratio as the
basic quantity for our finite-size scaling study. In particular, we
shall consider ratios of higher momenta:
\begin{equation}\label{eq:R12}
R_{12} = \frac{\hat G (\boldsymbol k_1)}{\hat G(\boldsymbol k_2)},
\end{equation}
where $\boldsymbol k_1=(2\pi/L,0,0,0)$, $\boldsymbol k_2 = (2\pi/L, 2\pi/L, 0,0)$ (and permutations)
are the smallest non-zero momenta compatible with the periodic 
boundary conditions.
Notice that, while our use of $R_{12}$ as a basic parameter is novel,
this is not in any way a strange quantity. In fact, it is a universal
renormalisation-group invariant, whose value in the large-$L$ limit
for a paramagnetic system should be $R_{12}(T>T_\text{c})=1$.  At the
critical point, however, $R_{12}(T_\text{c})>1$. For instance, using
conformal theory relations \cite{difrancesco:87,difrancesco:88}, we have
computed the critical ratio exactly for the non-disordered $D=2$ Ising model:
$R_{12}^\text{Ising}(T_\text{c}) = 1.694\,024\ldots$

To leading order, $R_{12}$ should have the same
scaling behaviour as $\xi_2/L$, namely,
\begin{equation}\label{eq:R12-FSS}
R_{12} = f_{12}\bigl( L^{1/\nu} t \bigr) + \text{[scaling corrections]}.
\end{equation}
However, since this quantity avoids the anomalous $\boldsymbol k=0$ mode, we
expect that corrections to scaling be smaller.
Indeed, in the bottom panel of Figure~\ref{fig:xi-R12} we can see
that the improvement in the scaling from the $\xi_2$ case is dramatic.
Even though corrections to scaling are noticeable,
for large sizes the intersections of the curves seem to
converge. Notice as well that the high values of $R_{12}$ in the
neighbourhood of the intersection point are not only far from the
paramagnetic limit of $R_{12}=1$, but also above the bound $R_{12}\leq
2$ that would result from a smooth behaviour of the 
propagator (see the discussion following eq.~\eqref{eq:OZ}).

Therefore, it is our working hypothesis that there is a phase
transition, but one that is affected by large corrections to scaling.
To substantiate this statement and actually compute the
critical parameters, we must begin by somehow controlling these
corrections. This analysis is rather technical, but not critical
to our discussion, so we leave it
for Appendix~\ref{sec:omega}, where we study the behaviour of $\xi_2/L$ at fixed $R_{12}$
as a function of $L$. To leading order, this should be a constant, 
so it has allowed us to isolate the effect of the scaling corrections, parameterised
as an extra term in $L^{-\omega}$ in~\eqref{eq:xi-FSS} and~\eqref{eq:R12-FSS}, where
\begin{align}\label{eq:omega}
\omega &= 1.43(37).
\end{align}

Now that we have the scaling corrections exponent $\omega$, we can go back to
study $R_{12}$. The easiest way to compute the critical parameters
($T_\text{c}$, $\nu$, etc.) from a renormalisation-group invariant such as
$R_{12}$ is the quotients method \cite{ballesteros:96}. Unfortunately, in our
case the corrections to scaling are strong, and for some lattice sizes we do
not actually reach the intersection point. Let us, therefore, consider an
alternative procedure \cite{billoire:11}. 

We assume (and therefore will test) that all points of the de Almeida-Thouless
line ($h>0$) belong to the same universality class.  We begin by considering
eq.~\eqref{eq:R12-FSS} and explicitely write the corrections to scaling,
recall that $t=(T-T_\mathrm{c}(h))/T_\mathrm{c}(h)$,
\begin{equation}
R_{12}\bigl(T, L,h\bigr) = f_{12}(t L^{1/\nu}) + A(h,t L^{1/\nu}) L^{-\omega} + \ldots
\end{equation}
Now we define $T_R^L(h)$ as 
\begin{equation}
R_{12}\bigl(T_R^L(h), L, h\bigr) = R.
\end{equation}
Therefore, if $R$ is in the scaling region, i.e., not too far from $f_{12}(0)$,
then 
\begin{equation}\label{eq:TyL}
T_R^L(h) \simeq T_\text{c}(h) + B_{R,h} L^{-1/\nu}[1+ C_{R,h} L^{-\omega}].
\end{equation}
\begin{figure}[t]
\centering
\includegraphics[height=\linewidth,angle=270]{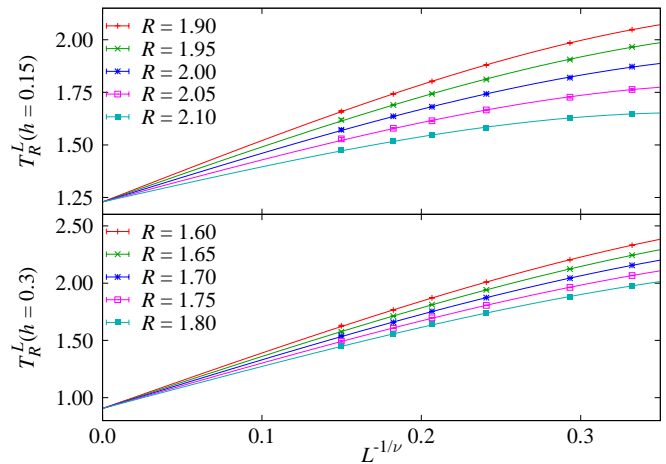}
\caption{Computation of the critical temperature $T_\text{c}(h)$ and
the critical exponent $\nu$. We compute the temperature $T_R^L(h)$ for
which $R_{12}(T_R^L,L,h)=R$. For $R$ inside the scaling region, these
temperatures should approach the critical one according
to~\eqref{eq:TyL}. We perform a joint fit for all the data sets in the
plot, forcing all of them to share the same $\nu$ and forcing all sets
with the same $h$ to extrapolate to the same $T_\text{c}(h)$. 
The result of this fit, which had a chi-square per degree of freedom 
of $\chi^2/\text{d.o.f.} = 40.2/37$ ($P$-value: 33\%), 
can be seen in Table~\ref{tab:parameters}.
\label{fig:nu}}
\end{figure}
  Using this formula, keeping $\omega$ fixed to the value
of~\eqref{eq:omega}, we can, in principle, estimate the critical exponent
$\nu$ and the critical temperature $T_\text{c}(h)$. However, for a single
value of $R$ we do not have enough degrees of freedom in the
fit. Therefore, following \cite{janus:10b}, we consider several values of
$R$ and two values of the field, $h=0.15$ and $h=0.30$ at the same time in
a joint fit, where $\nu$ is shared by all data sets and $T_\text{c}(h)$ is
shared by all the data sets with the same $h$ (see \cite{yllanes:11} for
full details on this fitting procedure). This is plotted in
Figure~\ref{fig:nu},  while the fit parameters can be seen in
Table~\ref{tab:parameters}. We also include the critical temperature for
$h=0.075$, extrapolated with the value of $\nu$ computed for $h=0.15$ and
$h=0.3$~\footnote{The data for $h=0.075$ presented very severe corrections,
  probably due to the proximity of the $h=0$ critical point. Therefore, 
  we only use the data for $L\geq12$ in order to estimate
  $T_\text{c}(h=0.075)$.}.  We have thus been able to obtain a precise
determination of $T_\text{c}(h)$ and of the critical exponent $\nu$.  It is
important to mention that the value of $\nu$ which, as we have seen, is
universal for $h>0$, is very different from that of the $h=0$ case.  As a
consistency check of our non-standard finite-size scaling method, we have
run a smaller set of simulations for $h=0$ and obtained $\nu^{(0)}=0.96(11)$
and $T_\mathrm{c}^{(0)}=2.002(10)$, in good agreement with previous results
for the $h=0$ case \cite{jorg:08c,marinari:99b}. We remark that both the
critical temperature and $\nu^{(0)}$ widely differ from the values in
Table~\ref{tab:parameters}.
\begin{table}[t]
\begin{ruledtabular}
\begin{tabular}{cccc}
 Parameter  & $h=0.3$ & $h=0.15$ & $h=0.075$\\
\hline
$T_\text{c}(h)$ & 0.906(40)[3] & 1.229(30)[2] &  1.50(7)  \\ 
$\nu$            & \multicolumn{2}{c}{1.46(7)[6]} & ---\\
$\eta$           & \multicolumn{2}{c}{$-0.30(4)[1]$} & ---\\
\end{tabular}
\end{ruledtabular}
\caption{Critical temperatures and exponents for our model.
The second error bar, in square brackets, refers to the effect
of the uncertainty in $\omega$.
In order to compute $\nu$ and $T_\text{c}(h)$ we studied 
the scaling of $R_{12}$ with a joint fit for $h=0.15$ and $h=0.3$, 
as depicted in Figure~\ref{fig:nu}. The value of $\eta$ was computed
from the scaling of $\hat G(\boldsymbol k)$ at $h=0.15$ and $h=0.3$.
The data for $h=0.075$  presented severe corrections, probably due to 
the proximity of the $h=0$ critical point. Therefore, we did not 
include this field in the previous fits and only used the data for $L\geq12$
in order to estimate $T_\text{c}(h=0.075)$ using the previously 
computed $\nu$. 
\label{tab:parameters} }
\end{table}

  The determination of the second independent
critical exponent, the anomalous dimension $\eta$ of the propagator, is much
more difficult.  In principle, we could consider the scaling of the
propagator $\hat G(\boldsymbol k)$ at fixed $R_{12}=R$. However, $\eta$ is
more affected by the scaling corrections than $\nu$. In fact, as discussed
in Appendix~\ref{sec:omega}, we had to consider quadratic corrections
to scaling, as $A_1L^{-\omega_\text{eff}} + A_2 L^{-2\omega_\text{eff}}$,
with $\omega_\text{eff}=2.2(3)$, in order to fit the data. Our final
estimate is quoted in Table~\ref{tab:parameters}.  Finally, we can combine
our results in Table~\ref{tab:parameters} to sketch the de Almeida-Thouless
line. This is plotted in the inset to Figure~\ref{fig:xi}, where we also
show a very good fit to the Fisher-Sompolinsky scaling \cite{fisher:85}. 

Let us finally mention that one may analyse our data as well under the
assumption of the {\em absence} of the phase transition in a field. This
analysis, which relies on Refs.~\cite{fisher:88,hartmann:99b,hukushima:99},
is reported in Appendix~\ref{sec:no-transition}. The data fail to
follow basic scaling relations derived under the no-transition hypothesis (or,
at least, they fail to scale within the range of system sizes that we could
simulate).

\section{Discussion}
In summary, we have presented a finite-size scaling study of the four
dimensional Edwards-Anderson model of an Ising spin glass in an external
magnetic field. We have been able to reach large system sizes and low
temperatures, thanks to the Janus special-purpose computer. We introduce a
novel finite-size scaling method, which cures the anomalies first observed in
\cite{leuzzi:09}. We present conclusive evidence for the presence
of a de Almeida-Thouless line in the temperature-magnetic field phase plane
(inset for Figure~\ref{fig:xi}), whose universality class we characterise. In
other words, a spin-glass transition occurs, even without time-reversal
symmetry, for realistic models (i.e., well below the upper critical dimension
$d_\mathrm{u}=6$). A far-reaching consequence is that the universality class
for the phase transition in structural glasses may actually
exist \cite{moore:02}. Our result also settles a longstanding controversy in
the field of spin glasses (see, e.g., \cite{bray:11,parisi:11}).


\begin{acknowledgments}
We thank Davide Rossetti for introducing us to the handling of the 128-bit SSE
registers. We acknowledge partial financial support from MICINN, Spain,
(contract nos. FIS2009-12648-C03, FIS2010-16587, TEC2010-19207), from UCM-Banco de Santander
(GR32/10-A/910383), from Junta de Extremadura, Spain (contract no. GR10158)
and from Universidad de Extremadura (contract no. ACCVII-08).  B.S. and
D.Y. were supported by the FPU program (Ministerio de Educaci\'on, Spain);
R.A.B. and J.M.-G. were supported by the FPI program (Diputaci\'on de
Arag\'on, Spain); finally J.M.G.-N. was supported by the FPI program
(Ministerio de Ciencia e Innovaci\'on, Spain).

\end{acknowledgments}
\appendix
\section{Simulations}\label{sec:simulation-details}
\begin{table}[t]\centering
\begin{ruledtabular}
 \begin{tabular}{clcccc}
$L$  & \multicolumn{1}{c}{$h$} & $T_{\mathrm{min}}$ & $T_{\mathrm{max}}$ & $N_{T}$ & $N_\text{s}$ \\
\hline
$5$  & $0.075$ & $1.300$  & $2.600$    & $14$  &  $25600$ \\
$5$  & $0.150$ & $1.300$  & $2.600$    & $14$ &  $25600$ \\
$5$  & $0.300$ & $0.833$  & $2.797$    & $20$ &  $25600$ \\
\hline
$6$  & $0.075$ & $1.300$  & $2.600$    & $14$ &  $25600$  \\
$6$  & $0.150$ & $1.300$  & $2.600$    & $14$  &  $25600$ \\
$6$  & $0.300$ & $0.833$  & $2.797$    & $20$  &  $25600$ \\
\hline
$8$  & $0.075$ & $1.350$  & $2.500$    & $24$  &  $25600$  \\
$8$  & $0.150$ & $1.350$  & $2.500$    & $24$  &  $25600$ \\
$8$  & $0.300$ & $1.200$  & $2.325$    & $16$  &  $25600$ \\
\hline
$10$ & $0.075$ & $1.350$ & $2.402$  & $26$    & $25600$ \\
$10$ & $0.150$ & $1.350$ & $2.402$  & $26$     & $25600$  \\
$10$ & $0.300$ & $1.340$ & $2.243$  & $20$      & $25600$ \\
\hline
$12$ & $0.075$ & $1.425$ & $2.402$  & $24$     & $25600$ \\
$12$ & $0.150$ & $1.502$ & $2.402$  & $22$    & $12800$ \\
$12$ & $0.3$ & $1.400$ & $2.160$  & $20$     & $25600$  \\
\hline
$16$ & $0.075$ & $1.400$ & $2.179$  & $32$  & $4000$ \\
$16$ & $0.15$ & $1.400$ & $2.179$  & $32$    & $4000$ \\
$16$ & $0.300$ & $1.304$ & $1.681$  & $18$     & $1000$ \\
 \end{tabular}
\end{ruledtabular}
\caption{Parameters describing our parallel tempering simulations. The $N_T$
  temperatures were evenly spaced between $T_\mathrm{min}$ and
  $T_\mathrm{max}$ for $L\le 12$, but with a slightly larger separation in the
  hot region for $L=16$. We simulate four real replicas for each of the
  $N_\mathrm{s}$ samples.
\label{tab:details} }
\end{table}
We have carried out parallel tempering \cite{hukushima:96,marinari:98b}
simulations for three values of the magnetic field (see Table~\ref{tab:details}), with periodic boundary conditions.

Our thermalisation protocol is sample dependent (see
\cite{yllanes:11} for details). We first perform a number of
iterations large enough to ensure thermalisation in a large fraction of the
samples (typically 90\%). We study the autocorrelation of the temperature flow
during the parallel tempering and extend the runs for the slower samples until
a total length of 14 exponential autocorrelation times is ensured. The final
product is a set of thermalised and almost independent configurations. As an
example, each $L=16$ sample in $h=0.15$ was simulated at least for $5\times
10^7$ heat bath lattice sweeps at each of the $N_T=32$ temperatures (we
performed a parallel tempering update every 10 heat baths). However, the
hardest sample required as many as $2.6\times 10^{10}$ heat bath sweeps.

The $L=16$ lattices were simulated on the Janus computer with an update speed
(for each of its 256 units) of $86$ ps per spin flip with a heat bath scheme.
The $L\le 12$ lattices were simulated on PC clusters, with a C code that uses
multi-spin coding \cite{newman:99} with 128-bit words (using the SSE extensions);
the update speed in this case is 350 ps per spin flip using a Metropolis
algorithm (on an Intel Core2 processor at 2.40GHz)
With multi-spin coding, the samples whose simulations have to be
extended must be extracted from the original 128-sample bundles to construct new
bundles that are then extended with the same code.
Note finally that, since the PC spreads the spin-flips over 128 samples, the
simulation for each sample is faster on Janus by a factor $\sim 500$. This
difference is significant when the equilibration time is large. 

\section{The correlation functions}\label{sec:correlation-functions}
The main quantities that we compute are the correlation functions. In the
presence of a magnetic field, the expectation of each spin
$S_{\boldsymbol{x}}$ is non-vanishing. Hence we may consider these two
correlation functions:
\begin{eqnarray}
G_1(\boldsymbol r)&=& \frac{1}{L^4} \sum_{\boldsymbol{x}} \overline{\bigl(\langle S_{\boldsymbol{x}} S_{\boldsymbol{x}+\boldsymbol r}\rangle -\langle S_{\boldsymbol{x}}\rangle\langle S_{\boldsymbol{x}+\boldsymbol r}\rangle\bigr)^2}\,,\\
G_2(\boldsymbol r)&=& \frac{1}{L^4} \sum_{\boldsymbol{x}} \overline{\bigl(\langle S_{\boldsymbol{x}} S_{\boldsymbol{x}+\boldsymbol r}\rangle^2 -\langle S_{\boldsymbol{x}}\rangle^2\langle S_{\boldsymbol{x}+\boldsymbol r}\rangle^2\bigr)}\,.
\end{eqnarray}
In the above, the $\langle\cdots\rangle$ stands for the thermal
average in a single sample, while the disorder average is indicated by
an overline. Note that the Fourier transform $\hat G_1(\boldsymbol k
=0)$ is the spin-glass susceptibility. We simulate four real
replicas $\{S_{\boldsymbol x}^{(a)}\}$ (i.e., four
systems with the same coupling evolving independently under the thermal
noise) in order to obtain unbiased estimators of the correlation
functions. In the main text $G$ stands for either of the $G_{1,2}$. In
the fits we have combined data from both whenever it was useful to
obtain smaller statistical errors.

The correlation functions were computed off-line over stored
configurations.  We note that configurations at different Monte Carlo
times can be combined as long as they belong to different
replicas \cite{janus:10}. This results in small Monte Carlo errors with
a modest number of configurations, so the uncertainty on the final
result is dominated by the sample-to-sample fluctuations.  This step
is rather time consuming, so we also use multi-spin coding to
accomplish it.

In order to define the second-moment correlation length \cite{cooper:82},
we consider the following Ornstein-Zernike expansion for the propagator 
in Fourier space,
\begin{equation}\label{eq:OZ}
\frac{1}{\hat G(\boldsymbol k)} = \frac{\xi^2}{\hat G(0)}\left[\frac{1}{\xi^2} + \cancel k^2 + a_4 (\cancel k^2)^2+\ldots\right],
\end{equation}
where $\cancel k^2=4\sum_{\mu} \sin^2(k_\mu/2)$.
Then, the common second-moment correlation length $\xi_2$ is 
obtained by truncating the expansion at the $\cancel k^2$ term:
\begin{equation}\label{eq:xi2}
\xi_2 = \frac{1}{2\sin(\pi/L)} \left(\frac{\hat G(0)}{\hat G(\boldsymbol k_1)} -1 \right)^{1/2},
\end{equation}

As we comment in the main text, this definition is not
well behaved for our model, due to the anomalous
behaviour of the $\boldsymbol k=0$ mode.  Actually, there is a simple,
yet unexpected explanation for this
anomaly. It arises
whenever soft excitations (Goldstone bosons) are present in the
low-temperature phase, while an external magnetic field splits excitations into
longitudinal and transversal \cite{zinn-justin:05}. Familiar examples of
Goldstone bosons are magnons, or the phonons in an acoustical
branch. What is most peculiar about spin glasses is that soft modes
are present \cite{bray:87,dedominicis:98}, even if our variables are
discrete.

\section{Computation of the scaling corrections}\label{sec:omega}
In order to compute corrections to scaling, we recall that the 
behavior of $\xi_2/L$ and $R_{12}$ is the same to leading scaling order:
\begin{align}\label{eq:sup1}
\xi_2/L(T,L,h) &= f_\xi(t L^{1/\nu}) + \text{[scaling corrections]},\\
R_{12}(T,L,h) &= f_{12}(t L^{1/\nu}) + \text{[scaling corrections]}.
\end{align}
As we saw in Figure~2, the effect of the scaling corrections was dramatic
for $\xi_2/L$ (they erase all trace of a phase transition in the simulated
regime). For $R_{12}$, the deviation from the leading scaling was milder, 
but still strong enough that we cannot ignore it in our analysis. Therefore, 
it becomes necessary to parameterize the corrections.

\begin{figure}
\centering
\includegraphics[height=\linewidth,angle=270]{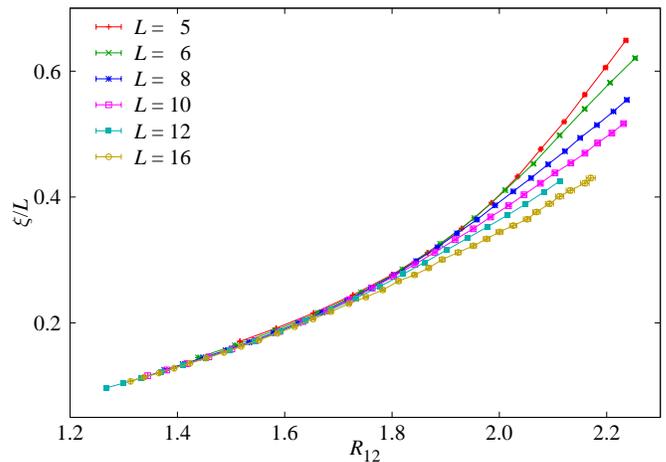}
\caption{Plot of $\xi_2/L$ 
against $R_{12}$ at $h=0.15$. To leading scaling order, this 
curve should be independent of the system size, a prediction that is realized 
at low values of $R_{12}$ (i.e., far from the critical point).
Close to the critical point, however, strong scaling corrections appear.
\label{fig:sup1}
}
\end{figure}

\begin{figure}
\centering
\includegraphics[height=\linewidth,angle=270]{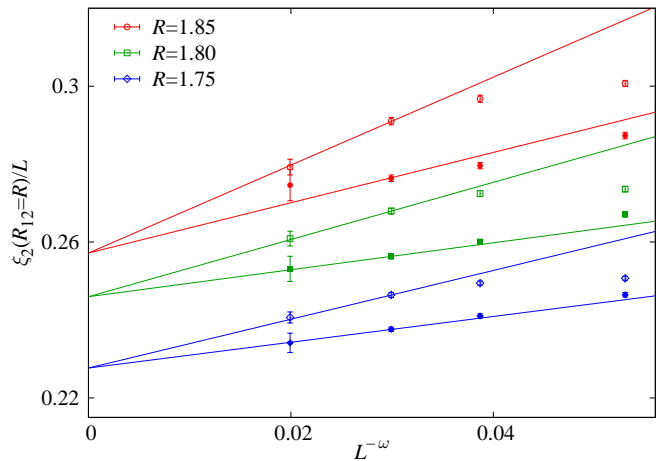}
\caption{Plots of $\xi_2/L$ at fixed $R_{12}=R$ as a 
function of the lattice size (i.e., vertical cuts of the left
panel) for $h=0.15$ (empty symbols) and $h=0.3$ (solid symbols).
The extrapolation to the $h$-independent
limit for each $R$ is governed by the scaling corrections exponent
$\omega$ ---eq.~\ref{eq:xi2-corrections} .  We have
performed a joint fit for several values of $R$, forcing all data sets
to share the same exponent and taking their correlation into account\label{fig:sup2}
}
\end{figure}

Since the number of lattice sizes is not enough to fit for the leading 
order behavior and the corrections at the same time, our strategy has
been isolating the latter. In order so to do, we first consider the 
evolution of $\xi_2/L$ as a function of $R_{12}$ (Figure~\ref{fig:sup1}).
According to~\eqref{eq:sup1}, to leading order this function should not depend
on the lattice size. We see in the figure that this behavior is indeed realized
at high temperatures (low values of $R_{12}$) but that, close to the transition 
point, the deviations become very strong. Therefore, by fixing the value
of $R_{12}$ and studying the evolution of $\xi_2/L$ with the system size
we can isolate the corrections to scaling. These include an enormous
variety of effects, but the standard practice is 
to consider simply the leading corrections, characterized by an exponent $\omega$
\begin{equation}\label{eq:xi2-corrections}
\begin{split}
\xi_2(T,L,h)/L \simeq &g\bigl(R_{12}(T,L,h)\bigr)\\& + A\bigl(R_{12}(T,L,h),h\bigr) L^{-\omega}.
\end{split}
\end{equation}
Notice that the extrapolation to infinite volume depends only on $R_{12}$, but 
that the amplitude of the corrections depends also separately on the 
magnetic field $h$. We can, therefore, consider a joint fit
to~\eqref{eq:xi2-corrections} for $h=0.15,0.3$ and several values 
of $R_{12}$. Data for $h=0.075$ are not included due to their proximity to $h=0$.
In this global fit, the value of $\omega$ is shared 
by all data sets, while the infinite-volume extrapolation is 
common to the data sets with the same $R_{12}$ but different $h$.
This is represented in Figure~\ref{fig:sup2}, where we only show a few
$R_{12}$ values for clarity. In this computation we use our data
for $L>10$. The result and
the chi-square per degree of freedom (as well as the $P$-value of the
fit) are
\begin{align}
\omega &= 1.43(37), & \chi^2/\text{d.o.f.} &= 9.2/11\  (P=60\%).
\end{align}
This is the value of $\omega$ used in the main text of the paper to
compute $\nu$ and the critical temperatures. Notice, however, 
that for its computation we had to restrict the fit to
lattice sizes $L\geq 12$. For smaller systems, the effect
of subleading corrections to scaling is very important, as 
evinced by the curvature in the plots of the figure.

Since this is a potentially
dangerous effect, we have also, as a consistency check, tried to include
smaller lattices in the fit at the cost of including more correction
terms. Unfortunately, these subleading corrections
are extremely varied and we cannot be sure of
the relative importance of each term. We do know, however, that among
them is a series on powers of $L^{-\omega}$. Therefore, we can 
consider an effective parameterization of the corrections in the following way
\begin{equation}
\begin{split}
\xi_2(T,L,h)/L \simeq& g\bigl(R_{12}(T,L,h)\bigr)\\& + B\bigl(R_{12}(T,L,h),h\bigr) L^{-\omega_\text{eff}}\\ &
+C\bigl(R_{12}(T,L,h),h\bigr) L^{-2\omega_\text{eff}}.
\end{split}
\end{equation}
If we redo the computation of Figure~\ref{fig:sup2} in
this non-standard way, including lattices with $L>6$ we obtain a value of
$\omega_\text{eff}=2.2(3)$. At a first glance the discrepancy
with the previous $\omega$ may seem alarming. Actually, 
a computation of $\nu$ and the critical temperatures with this
alternative corrections to scaling yields 
compatible values: $T_\text{c}(0.3) = 0.902(33)[1],$
$T_\text{c}(0.15)=1.233(23)[1]$ and $\nu=1.54(6)[2]$, with
$\chi^2/\text{d.o.f.}=39.0/37$ $(P=38\%)$.  This can be seen as a
check that, even though the scaling corrections are strong, our
computation of the critical parameters is robust.

A final comment regards the estimation of the second independent critical
exponent, $\eta$. We can consider the scaling of the propagator $\hat G(\boldsymbol k)$
at fixed $R_{12}=y$ (i.e., at $T=T_y^L$, see main text), with either of the 
two models for the scaling corrections:
\begin{align}
\hat G(\boldsymbol k, T_y^L, L, h)&\simeq  A_{k,h,y} L^{2-\eta}[1+B_{k,h,y}L^{-\omega}].\label{eq:eta1}\\
\hat G(\boldsymbol k, T_y^L, L, h)& \simeq  A_{k,h,y} L^{2-\eta}[1+C_{k,h,y}L^{-\omega_\text{eff}} \\& \hspace*{2.85cm} D_{k,h,y}L^{-2\omega_\text{eff}}]. \nonumber \label{eq:eta2}
\end{align}
In this case we have found that the first form, including only the leading corrections, 
does not adequately represent our data unless
we exclude from the fit so many lattice sizes that we lose all degrees
of freedom. Therefore, we cannot make a controlled determination of
this critical exponent using~\eqref{eq:eta1}.  The most that can be said is that the
propagator $\hat G(\boldsymbol k)$ ---and, in particular, the
susceptibility $\tilde \chi=\hat G(0)$--- diverges quickly at the
critical point, with a value of $\eta\approx-0.3$, similar to the
value for $h=0$.

On the other hand, using the effective quadratic corrections of~\eqref{eq:eta2},
the value of this exponent results
\begin{align}
\eta &= -0.30(4)[1],
\end{align}
with $\chi^2/\text{d.o.f.} = 9.55/11$  ($P=57\%$).
This is the value quoted in Table~\ref{tab:parameters} of the paper.

\section{The no-transition hypothesis}\label{sec:no-transition}

Let us assume that no real phase transition arises in a field.  For the sake
of brevity, we introduce the reduced temperature, which depends on the
standard temperature $T$:
\begin{equation}\label{eq:t-def}
t=\frac{T-T_\mathrm{c}}{T_\mathrm{c}}\,.
\end{equation}
In particular, note that $t(T_\mathrm{c})\!=\!0$, while $t(T\!=\!0)\!=\!-1$.
In this section, we are assuming that the only phase transition occurs when
$h=0$, hence $T_\mathrm{c}$ refers to the zero-field critical temperature.
Therefore, in a field, the correlation length in the thermodynamic limit is
finite for all temperatures: $\xi(t,h)<\infty$ if $h>0$. Under these
circumstances, it is unavoidable that
\begin{equation}\label{eq:limit}
\lim_{L\to\infty} R_{12}(t,h,L)=1\,.
\end{equation}
Nevertheless, it is quite reasonable to expect that $R_{12}$ verify a scaling
law ($W$ is a scaling function):
\begin{equation}\label{eq:basico}
R_{12}(t,h,L)=W(\xi(t,h)/L)
\end{equation}
In order to model our data according to this expectation, we need some
educated guess for $\xi(t,h)$. We shall get it from the droplet model
for spin glasses~\cite{bray:87,fisher:88}.

We first gather some necessary information in Sects.~\ref{sec:droplets}
and~\ref{sec:Tc}. Our data are analyzed under this light in
Sect.~\ref{sec:analysis}.

\subsection{The critical behavior according to the droplet model}\label{sec:droplets}

In the droplet model for spin glasses~\cite{fisher:88}, one
expects for $t=-1$ (i.e. $T=0$) a correlation length that diverges only in the
limit of $h\to 0$:
\begin{equation}\label{eq:droplet-T0}
\xi(t=-1,h)\propto h^{-2/(D-2\theta)}\,,
\end{equation}
where $\theta$ is the droplet exponent. In $D=4,$ that exponent is
$\theta\approx0.7$~\cite{hartmann:99b,hukushima:99}.
So, the prediction is 
\begin{equation}\label{eq:droplet-T0-numerico}
\xi_\mathrm{droplet}(t=-1,h)\propto 1/h^{0.77}\,.
\end{equation}

Fisher and Huse~\cite{fisher:88}, define as well a {\em
  dynamical} de Almeida-Thouless (dAT) line. That is, a freezing temperature
which should scale very much as the equilibrium dAT line (which is inexistent
on their theory). This freezing line $T_f(t_m,h)$ would depend on the
measuring time $t_m$ as ($\tau_0$ is a microscopic time unit and $\Psi$ is the
barrier exponent)
\begin{equation}\label{eq:dinamica-droplet}
\frac{T_\mathrm{c}- T_f(t_m,h)}{T_\mathrm{c}} \sim  h^{\frac{2}{\gamma+\beta}} 
[\log (t_m/\tau_0)]^{(D-\theta)/(\gamma+\beta)\Psi}\,. 
\end{equation}
Hence, for a fixed measuring time $t_m$, the freezing line scales with $h$
just as expected for the dAT line, see Sect.~\ref{sec:Tc} below.

Fisher and Huse description~\cite{fisher:88} of the crossover phenomena to be
observed in equilibrium, coincides with our Eq.~\eqref{eq:xi-ansatz} (see
below).

\subsection{Scaling close to the $h=0$ critical point}\label{sec:Tc}

The critical behavior of the Edwards-Anderson model, in $D=4$ and $h=0$, is
relatively well understood~\cite{jorg:08c,marinari:99b}:
\begin{align}
T_\mathrm{c}=2.03(3)\,,\\ 
\nu =1.025(15)\,,\\
\eta=-0.275(25)\,.
\end{align}
From these estimates, we obtain the
Renormalization Group (RG)  eigenvalues $y_t$ and $y_h$,
\begin{align}
y_t&=1/\nu=0.976(14)\,,\\
y_h&=\frac{D+2-\eta}{2}=3.137(13)\,.
\end{align}
In order to make connection with Eq.~\eqref{eq:dinamica-droplet}, note, in
particular, that $y_t/y_h=1/(\beta+\gamma)$.

The Fisher-Sompolinski relation~\cite{fisher:85}, follows from a
simple scaling argument.  Recall that, for spin glasses, the ordering field is
$h^2$ rather than $h$, due to the gauge invariance of the coupling
distribution. Hence, the RG transformation of scale $b$ transforms $h^2$ into
$b^{y_h} h^2$, while the correlation length transforms as (see, e.g., \cite{amit:05})
\begin{equation}\label{eq:simple-scaling}
\xi(t,h)= b\, \xi(b^{y_t} t, b^{y_h/2} h)\,.
\end{equation}
Let us now choose $b$ such that  $b^{y_h} h^2 =  1$, hence
\begin{equation}\label{eq:Fisher-Sompolinski}
\xi(t,h^2)= \frac{1}{h^{2/y_h}} G (t/h^{2 y_t/y_h})\,,
\end{equation}
where $G(x)$ is a scaling function.  Now, imagine that there {\em is} a dAT
line. Then, one should be able of adjusting the temperature, at fixed $h>0$,
in such a way that $\xi(t,h)$ grows unboundly. On the view of
Eq.~\eqref{eq:Fisher-Sompolinski}, this is only possible if the scaling
function $G(x)$ has a singularity at some $x^*$. Hence, the dAT line would be
located at
\begin{equation}
t=x^* h^{2 y_t/y_h}= x^* h^{2/(\beta+\gamma)}\,,
\end{equation}
at least while the scaling fields $t$ and $h$ are small enough to behave
linearly under the RG, as we have assumed in Eq.~\eqref{eq:simple-scaling}.

\begin{figure}
\centering
\includegraphics[height=\linewidth,angle=270]{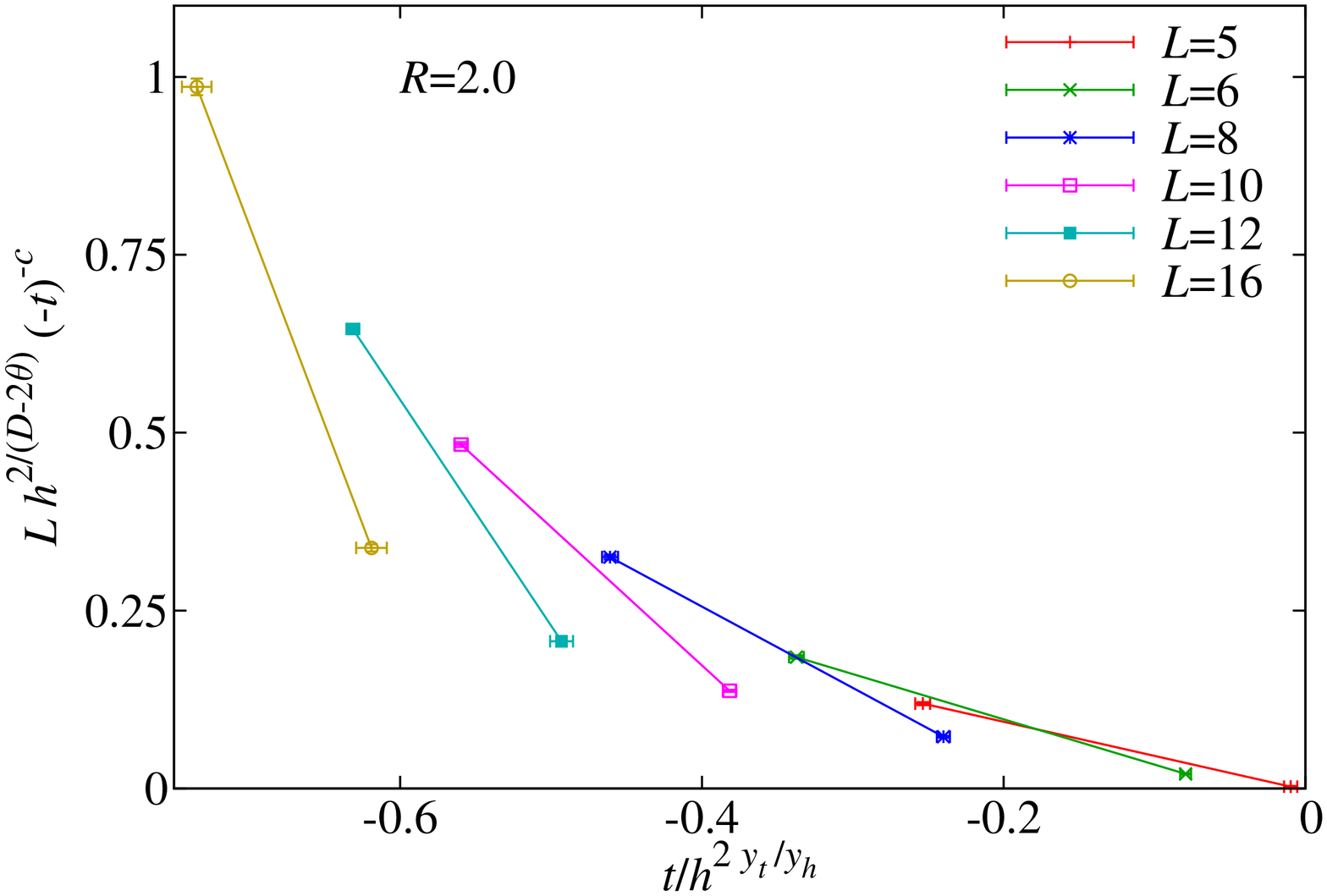}
\includegraphics[height=\linewidth,angle=270]{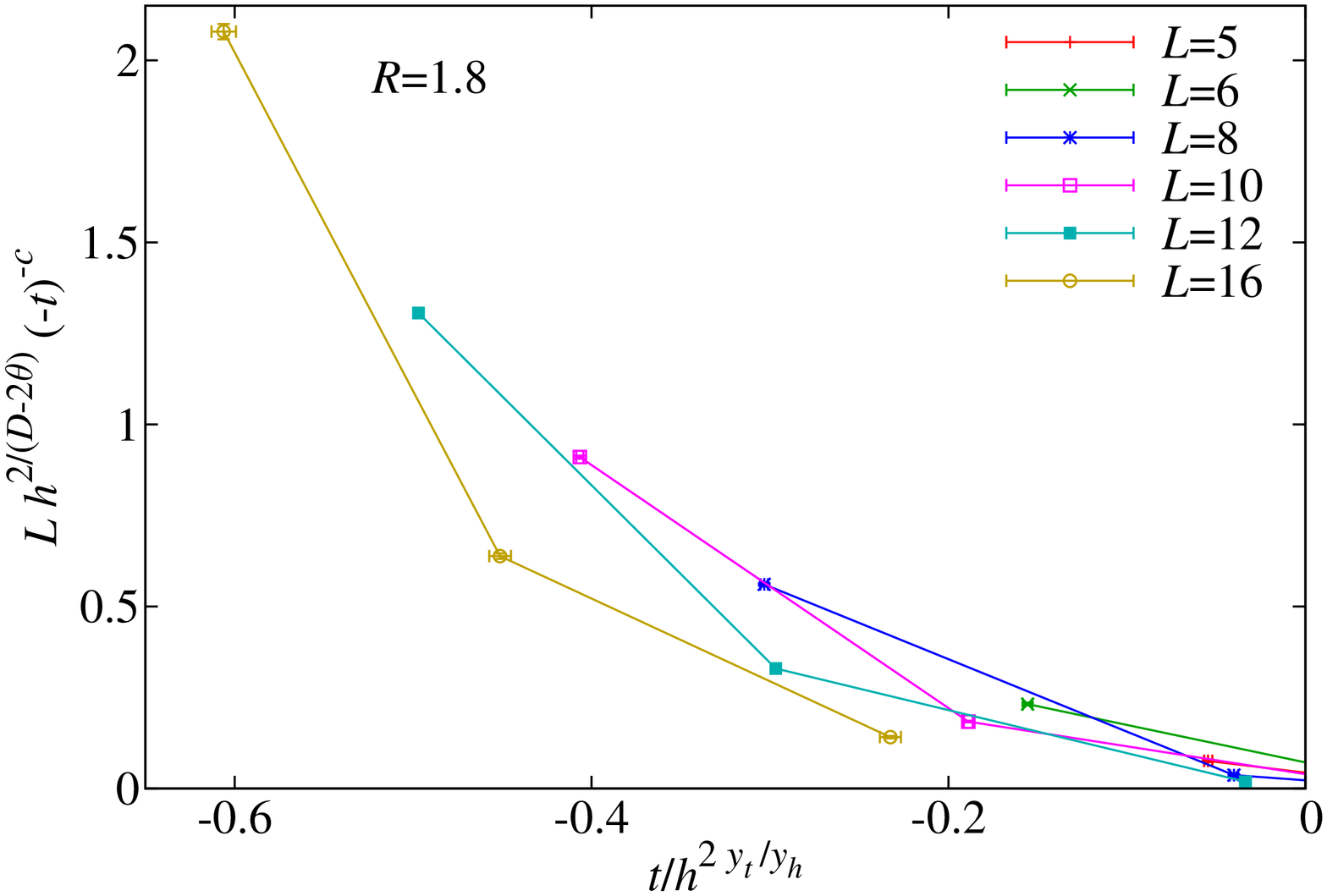}
\caption{Numerical test of Eq.~\eqref{eq:droplet-comprobar}. For every lattice
  size $L$, and fields $h$, we obtain the temperature where
  $R_{12}(T_R^L,L,h)=R$. The reduced temperature $t_R^L(h)$ is computed
  through Eq.~\eqref{eq:t-def} ($T_\mathrm{c}$ is the zero-field critical
  temperature).  According to the droplet model, and neglecting scaling
  corrections, for fixed $R$, $L h^{2/(D-2\theta)} [-t_R^L(h)]^c$ must be an
  $L$-independent function of $t_R^L(h)/h ^{2 y_t/y_h}$. We plot data for
  $R=2$ (top), and $R=1.8$ (bottom). For $R=2.0$, $T_R^L$ lies within our
  simulated temperature trange only for $h=0.15,0.075$. In the plot, we used
  $\theta=0.7$, which is an average among $\theta=0.65$ (from~\cite{hartmann:99b}),
  and $\theta=0.82$ (from~\cite{hukushima:99}).
 No matter the value that we take for $\theta$
  in this range, $L h^{2/(D-2\theta)} [-t_R^L(h)]^c$ grows quickly with $L$
  and draws an increasingly steep curve.
\label{fig:sup3}}
\end{figure}

\subsection{The ansatz for $\xi(t,h)$ and the comparison with numerical
  data}\label{sec:analysis}

As said above, we need some educated guess about the behavior of $\xi(T,h)$
in the droplet picture. We take inspiration from
Eqs.~\eqref{eq:Fisher-Sompolinski} and~\eqref{eq:droplet-T0}.

Of course, the scaling function $G(x)$ in Eq.~\eqref{eq:Fisher-Sompolinski}
must be regular for all $x$ because, according to the droplet picture, there is
no phase transition. However, the function $G(x)$ must be singular for $x\to
-\infty$ (which corresponds to taking first the limit $T\to 0$, i.e. $t=-1$,
and later the limit $h\to 0$). In fact, if $G(x)$ would tend to a constant for
large, negative $x$, one would have $\xi(T=0,h)\sim 1/h^{2/y_h}\approx
1/h^{0.64}$. This is inconsistent with Eq.~\eqref{eq:droplet-T0-numerico}
(i.e. $\xi\propto 1/h^{0.77}$).  Nevertheless, a mild singularity $G(x\to
-\infty) \sim (-x)^c$, where exponent $c$ verifies
\begin{equation}
\frac{2}{y_h} (y_t c +1)= \frac{2}{D-2\theta}
\end{equation}
would make equation~\eqref{eq:Fisher-Sompolinski} consistent
with Eq.~\eqref{eq:droplet-T0}. Solving for $c$, we get
\begin{equation}
c=\frac{y_h-D+2\theta}{y_t(D-2\theta)}\approx 0.21
\end{equation}
Hence, our ansatz for $\xi(t,h)$, inspired by the droplet theory,  is 
\begin{equation}\label{eq:xi-ansatz} \xi_\mathrm{ansatz}(t,h)=
  \frac{(-t)^c}{h^\frac{2}{D-2\theta}} F(t/h^{2 y_t/y_h})\,.
\end{equation}
The new scaling function is $F(x)=G(x)/(-x)^c$. $F(x)$ should remain finite in
the limit $x\to -\infty$ (note, however, that $F(x)\sim 1/(-x)^c$ for
$x\approx 0$, which corresponds to the neighborhood of $T_\mathrm{c}$).

Plugging $\xi_\mathrm{ansatz}(t,h)$ in~\eqref{eq:basico}, we get
\begin{equation}
\begin{split}
R_{12}(t,h,L) &= W(\xi_\mathrm{ansatz}(t,h)/L)\\
&= W\left(\frac{(-t)^c}{L
  h^{2/(D-2\theta)}} F(t/h^{2 y_t/y_h})\right)\,.\label{eq:final-droplet}\end{split}
\end{equation}

Now, in the main text we paid great attention to $T_R^L(h)$, namely the
temperature where $R_{12}(T_R^L,L,h)=R$ (one may compute easily the reduced
temperature $t_R^L(h)$). We thus recast Eq.~\eqref{eq:final-droplet} in a form
directly amenable to a scaling analysis:
\begin{equation}
L h^{2/(D-2\theta)} [-t_R^L(h)]^c W^{-1}(R) =
F\left(\frac{t_R^L(h)}{h ^{2 y_t/y_h}}\right)\,,\label{eq:droplet-comprobar}
\end{equation}
where $W^{-1}(x)$ is the inverse function of $W(x)$. 

Hence, the prediction of the droplet model is fairly simple. For fixed $R$,
and barring scaling corrections (negligible for large $L$), the numerical
estimate of the l.h.s., namely $L h^{2/(D-2\theta)} [-t_R^L(h)]^c$, must be an
$L$-independent function of $t_R^L(h)/h ^{2 y_t/y_h}$. On the other hand, if a
phase transition occurs at fixed $t(h)/h ^{2 y_t/y_h}$ (as claimed in the main
text), $t_R^L(h)$ should tend to $t(h)$ while the l.h.s. should diverge as $L$
grows. In other words, if there is a dAT line, data should tend to a vertical
line in the large $L$ limit.

These two alternatives are compared in Figure~\ref{fig:sup3}.
In fact,
our estimate for $L h^{2/(D-2\theta)} [-t_R^L(h)]^c$ grows fast with $L$ and
draws an increasingly steep curve, which is the behavior expected in the
presence of a dAT line. This is hardly surprising, because the numerical data
in Figure~\ref{fig:xi-R12}--bottom, where $R_{12}$ shows scale invariant
behavior, are plainly inconsistent with our starting assumption in
Eq.~\eqref{eq:limit}.

In summary, either our simulated sizes are entirely in a preasymptotic
regime, or the no-transition hypothesis is not realized.


\begin{thebibliography}{47}%
\makeatletter
\providecommand \@ifxundefined [1]{%
 \@ifx{#1\undefined}
}%
\providecommand \@ifnum [1]{%
 \ifnum #1\expandafter \@firstoftwo
 \else \expandafter \@secondoftwo
 \fi
}%
\providecommand \@ifx [1]{%
 \ifx #1\expandafter \@firstoftwo
 \else \expandafter \@secondoftwo
 \fi
}%
\providecommand \natexlab [1]{#1}%
\providecommand \enquote  [1]{``#1''}%
\providecommand \bibnamefont  [1]{#1}%
\providecommand \bibfnamefont [1]{#1}%
\providecommand \citenamefont [1]{#1}%
\providecommand \href@noop [0]{\@secondoftwo}%
\providecommand \href [0]{\begingroup \@sanitize@url \@href}%
\providecommand \@href[1]{\@@startlink{#1}\@@href}%
\providecommand \@@href[1]{\endgroup#1\@@endlink}%
\providecommand \@sanitize@url [0]{\catcode `\\12\catcode `\$12\catcode
  `\&12\catcode `\#12\catcode `\^12\catcode `\_12\catcode `\%12\relax}%
\providecommand \@@startlink[1]{}%
\providecommand \@@endlink[0]{}%
\providecommand \url  [0]{\begingroup\@sanitize@url \@url }%
\providecommand \@url [1]{\endgroup\@href {#1}{\urlprefix }}%
\providecommand \urlprefix  [0]{URL }%
\providecommand \Eprint [0]{\href }%
\providecommand \doibase [0]{http://dx.doi.org/}%
\providecommand \selectlanguage [0]{\@gobble}%
\providecommand \bibinfo  [0]{\@secondoftwo}%
\providecommand \bibfield  [0]{\@secondoftwo}%
\providecommand \translation [1]{[#1]}%
\providecommand \BibitemOpen [0]{}%
\providecommand \bibitemStop [0]{}%
\providecommand \bibitemNoStop [0]{.\EOS\space}%
\providecommand \EOS [0]{\spacefactor3000\relax}%
\providecommand \BibitemShut  [1]{\csname bibitem#1\endcsname}%
\let\auto@bib@innerbib\@empty
\bibitem [{\citenamefont {Debenedetti}(1997)}]{debenedetti:97}%
  \BibitemOpen
  \bibfield  {author} {\bibinfo {author} {\bibfnamefont {P.~G.}\ \bibnamefont
  {Debenedetti}},\ }\href@noop {} {\emph {\bibinfo {title} {Metastable
  Liquids}}}\ (\bibinfo  {publisher} {Princeton University Press},\ \bibinfo
  {address} {Princeton},\ \bibinfo {year} {1997})\BibitemShut {NoStop}%
\bibitem [{\citenamefont {Debenedetti}\ and\ \citenamefont
  {Stillinger}(2001)}]{debenedetti:01}%
  \BibitemOpen
  \bibfield  {author} {\bibinfo {author} {\bibfnamefont {P.~G.}\ \bibnamefont
  {Debenedetti}}\ and\ \bibinfo {author} {\bibfnamefont {F.~H.}\ \bibnamefont
  {Stillinger}},\ }\href@noop {} {\bibfield  {journal} {\bibinfo  {journal}
  {Nature}\ }\textbf {\bibinfo {volume} {410}},\ \bibinfo {pages} {259}
  (\bibinfo {year} {2001})}\BibitemShut {NoStop}%
\bibitem [{\citenamefont {Cavagna}(2009)}]{cavagna:09}%
  \BibitemOpen
  \bibfield  {author} {\bibinfo {author} {\bibfnamefont {A.}~\bibnamefont
  {Cavagna}},\ }\href@noop {} {\bibfield  {journal} {\bibinfo  {journal}
  {Physics Reports}\ }\textbf {\bibinfo {volume} {476}},\ \bibinfo {pages} {51}
  (\bibinfo {year} {2009})},\ \Eprint {http://arxiv.org/abs/arXiv:0903.4264}
  {arXiv:0903.4264} \BibitemShut {NoStop}%
\bibitem [{\citenamefont {Mydosh}(1993)}]{mydosh:93}%
  \BibitemOpen
  \bibfield  {author} {\bibinfo {author} {\bibfnamefont {J.~A.}\ \bibnamefont
  {Mydosh}},\ }\href@noop {} {\emph {\bibinfo {title} {Spin Glasses: an
  Experimental Introduction}}}\ (\bibinfo  {publisher} {Taylor and Francis},\
  \bibinfo {address} {London},\ \bibinfo {year} {1993})\BibitemShut {NoStop}%
\bibitem [{\citenamefont {H{\'e}risson}\ and\ \citenamefont
  {Ocio}(2002)}]{herisson:02}%
  \BibitemOpen
  \bibfield  {author} {\bibinfo {author} {\bibfnamefont {D.}~\bibnamefont
  {H{\'e}risson}}\ and\ \bibinfo {author} {\bibfnamefont {M.}~\bibnamefont
  {Ocio}},\ }\href@noop {} {\bibfield  {journal} {\bibinfo  {journal} {Phys.
  Rev. Lett.}\ }\textbf {\bibinfo {volume} {88}},\ \bibinfo {pages} {257202}
  (\bibinfo {year} {2002})},\ \Eprint
  {http://arxiv.org/abs/arXiv:cond-mat/0112378} {arXiv:cond-mat/0112378}
  \BibitemShut {NoStop}%
\bibitem [{\citenamefont {Cruz}\ \emph {et~al.}(2001)\citenamefont {Cruz},
  \citenamefont {Pech}, \citenamefont {Tarancon}, \citenamefont {Tellez},
  \citenamefont {Ullod},\ and\ \citenamefont {Ungil}}]{cruz:01}%
  \BibitemOpen
  \bibfield  {author} {\bibinfo {author} {\bibfnamefont {A.}~\bibnamefont
  {Cruz}}, \bibinfo {author} {\bibfnamefont {J.}~\bibnamefont {Pech}}, \bibinfo
  {author} {\bibfnamefont {A.}~\bibnamefont {Tarancon}}, \bibinfo {author}
  {\bibfnamefont {P.}~\bibnamefont {Tellez}}, \bibinfo {author} {\bibfnamefont
  {C.~L.}\ \bibnamefont {Ullod}}, \ and\ \bibinfo {author} {\bibfnamefont
  {C.}~\bibnamefont {Ungil}},\ }\href@noop {} {\bibfield  {journal} {\bibinfo
  {journal} {Comp. Phys. Comm}\ }\textbf {\bibinfo {volume} {133}},\ \bibinfo
  {pages} {165} (\bibinfo {year} {2001})},\ \Eprint
  {http://arxiv.org/abs/arXiv:cond-mat/0004080} {arXiv:cond-mat/0004080}
  \BibitemShut {NoStop}%
\bibitem [{\citenamefont {Ogielski}(1985)}]{ogielski:85}%
  \BibitemOpen
  \bibfield  {author} {\bibinfo {author} {\bibfnamefont {A.}~\bibnamefont
  {Ogielski}},\ }\href@noop {} {\bibfield  {journal} {\bibinfo  {journal}
  {Phys. Rev. B}\ }\textbf {\bibinfo {volume} {32}},\ \bibinfo {pages} {7384}
  (\bibinfo {year} {1985})}\BibitemShut {NoStop}%
\bibitem [{\citenamefont {Belletti}\ \emph
  {et~al.}(2008{\natexlab{a}})\citenamefont {Belletti}, \citenamefont
  {Cotallo}, \citenamefont {Cruz}, \citenamefont {Fernandez}, \citenamefont
  {Gordillo}, \citenamefont {Maiorano}, \citenamefont {Mantovani},
  \citenamefont {Marinari}, \citenamefont {Martin-Mayor}, \citenamefont
  {Monforte}, \citenamefont {Mu{\~n}oz~Sudupe}, \citenamefont {Navarro},
  \citenamefont {Perez-Gaviro}, \citenamefont {Ruiz-Lorenzo}, \citenamefont
  {Schifano}, \citenamefont {Sciretti}, \citenamefont {Tarancon}, \citenamefont
  {Tripiccione},\ and\ \citenamefont {Velasco}}]{janus:08}%
  \BibitemOpen
  \bibfield  {author} {\bibinfo {author} {\bibfnamefont {F.}~\bibnamefont
  {Belletti}}, \bibinfo {author} {\bibfnamefont {M.}~\bibnamefont {Cotallo}},
  \bibinfo {author} {\bibfnamefont {A.}~\bibnamefont {Cruz}}, \bibinfo {author}
  {\bibfnamefont {L.~A.}\ \bibnamefont {Fernandez}}, \bibinfo {author}
  {\bibfnamefont {A.}~\bibnamefont {Gordillo}}, \bibinfo {author}
  {\bibfnamefont {A.}~\bibnamefont {Maiorano}}, \bibinfo {author}
  {\bibfnamefont {F.}~\bibnamefont {Mantovani}}, \bibinfo {author}
  {\bibfnamefont {E.}~\bibnamefont {Marinari}}, \bibinfo {author}
  {\bibfnamefont {V.}~\bibnamefont {Martin-Mayor}}, \bibinfo {author}
  {\bibfnamefont {J.}~\bibnamefont {Monforte}}, \bibinfo {author}
  {\bibfnamefont {A.}~\bibnamefont {Mu{\~n}oz~Sudupe}}, \bibinfo {author}
  {\bibfnamefont {D.}~\bibnamefont {Navarro}}, \bibinfo {author} {\bibfnamefont
  {S.}~\bibnamefont {Perez-Gaviro}}, \bibinfo {author} {\bibfnamefont {J.~J.}\
  \bibnamefont {Ruiz-Lorenzo}}, \bibinfo {author} {\bibfnamefont {S.~F.}\
  \bibnamefont {Schifano}}, \bibinfo {author} {\bibfnamefont {D.}~\bibnamefont
  {Sciretti}}, \bibinfo {author} {\bibfnamefont {A.}~\bibnamefont {Tarancon}},
  \bibinfo {author} {\bibfnamefont {R.}~\bibnamefont {Tripiccione}}, \ and\
  \bibinfo {author} {\bibfnamefont {J.~L.}\ \bibnamefont {Velasco}} (\bibinfo
  {collaboration} {Janus Collaboration}),\ }\href@noop {} {\bibfield  {journal}
  {\bibinfo  {journal} {Comp. Phys. Comm.}\ }\textbf {\bibinfo {volume}
  {178}},\ \bibinfo {pages} {208} (\bibinfo {year} {2008}{\natexlab{a}})},\
  \Eprint {http://arxiv.org/abs/arXiv:0704.3573} {arXiv:0704.3573} \BibitemShut
  {NoStop}%
\bibitem [{\citenamefont {Belletti}\ \emph
  {et~al.}(2008{\natexlab{b}})\citenamefont {Belletti}, \citenamefont
  {Cotallo}, \citenamefont {Cruz}, \citenamefont {Fernandez}, \citenamefont
  {Gordillo-Guerrero}, \citenamefont {Guidetti}, \citenamefont {Maiorano},
  \citenamefont {Mantovani}, \citenamefont {Marinari}, \citenamefont
  {Martin-Mayor}, \citenamefont {Mu{\~n}oz~Sudupe}, \citenamefont {Navarro},
  \citenamefont {Parisi}, \citenamefont {Perez-Gaviro}, \citenamefont
  {Ruiz-Lorenzo}, \citenamefont {Schifano}, \citenamefont {Sciretti},
  \citenamefont {Tarancon}, \citenamefont {Tripiccione}, \citenamefont
  {Velasco},\ and\ \citenamefont {Yllanes}}]{janus:08b}%
  \BibitemOpen
  \bibfield  {author} {\bibinfo {author} {\bibfnamefont {F.}~\bibnamefont
  {Belletti}}, \bibinfo {author} {\bibfnamefont {M.}~\bibnamefont {Cotallo}},
  \bibinfo {author} {\bibfnamefont {A.}~\bibnamefont {Cruz}}, \bibinfo {author}
  {\bibfnamefont {L.~A.}\ \bibnamefont {Fernandez}}, \bibinfo {author}
  {\bibfnamefont {A.}~\bibnamefont {Gordillo-Guerrero}}, \bibinfo {author}
  {\bibfnamefont {M.}~\bibnamefont {Guidetti}}, \bibinfo {author}
  {\bibfnamefont {A.}~\bibnamefont {Maiorano}}, \bibinfo {author}
  {\bibfnamefont {F.}~\bibnamefont {Mantovani}}, \bibinfo {author}
  {\bibfnamefont {E.}~\bibnamefont {Marinari}}, \bibinfo {author}
  {\bibfnamefont {V.}~\bibnamefont {Martin-Mayor}}, \bibinfo {author}
  {\bibfnamefont {A.}~\bibnamefont {Mu{\~n}oz~Sudupe}}, \bibinfo {author}
  {\bibfnamefont {D.}~\bibnamefont {Navarro}}, \bibinfo {author} {\bibfnamefont
  {G.}~\bibnamefont {Parisi}}, \bibinfo {author} {\bibfnamefont
  {S.}~\bibnamefont {Perez-Gaviro}}, \bibinfo {author} {\bibfnamefont {J.~J.}\
  \bibnamefont {Ruiz-Lorenzo}}, \bibinfo {author} {\bibfnamefont {S.~F.}\
  \bibnamefont {Schifano}}, \bibinfo {author} {\bibfnamefont {D.}~\bibnamefont
  {Sciretti}}, \bibinfo {author} {\bibfnamefont {A.}~\bibnamefont {Tarancon}},
  \bibinfo {author} {\bibfnamefont {R.}~\bibnamefont {Tripiccione}}, \bibinfo
  {author} {\bibfnamefont {J.~L.}\ \bibnamefont {Velasco}}, \ and\ \bibinfo
  {author} {\bibfnamefont {D.}~\bibnamefont {Yllanes}} (\bibinfo
  {collaboration} {Janus Collaboration}),\ }\href@noop {} {\bibfield  {journal}
  {\bibinfo  {journal} {Phys. Rev. Lett.}\ }\textbf {\bibinfo {volume} {101}},\
  \bibinfo {pages} {157201} (\bibinfo {year} {2008}{\natexlab{b}})},\ \Eprint
  {http://arxiv.org/abs/arXiv:0804.1471} {arXiv:0804.1471} \BibitemShut
  {NoStop}%
\bibitem [{\citenamefont {Gunnarsson}\ \emph {et~al.}(1991)\citenamefont
  {Gunnarsson}, \citenamefont {Svendlindh}, \citenamefont {Nordblad},
  \citenamefont {Lundgren}, \citenamefont {Aruga},\ and\ \citenamefont
  {Ito}}]{gunnarsson:91}%
  \BibitemOpen
  \bibfield  {author} {\bibinfo {author} {\bibfnamefont {K.}~\bibnamefont
  {Gunnarsson}}, \bibinfo {author} {\bibfnamefont {P.}~\bibnamefont
  {Svendlindh}}, \bibinfo {author} {\bibfnamefont {P.}~\bibnamefont
  {Nordblad}}, \bibinfo {author} {\bibfnamefont {L.}~\bibnamefont {Lundgren}},
  \bibinfo {author} {\bibfnamefont {H.}~\bibnamefont {Aruga}}, \ and\ \bibinfo
  {author} {\bibfnamefont {A.}~\bibnamefont {Ito}},\ }\href@noop {} {\bibfield
  {journal} {\bibinfo  {journal} {Phys. Rev. B}\ }\textbf {\bibinfo {volume}
  {43}},\ \bibinfo {pages} {8199} (\bibinfo {year} {1991})}\BibitemShut
  {NoStop}%
\bibitem [{\citenamefont {Ballesteros}\ \emph {et~al.}(2000)\citenamefont
  {Ballesteros}, \citenamefont {Cruz}, \citenamefont {Fernandez}, \citenamefont
  {Martin-Mayor}, \citenamefont {Pech}, \citenamefont {Ruiz-Lorenzo},
  \citenamefont {Tarancon}, \citenamefont {Tellez}, \citenamefont {Ullod},\
  and\ \citenamefont {Ungil}}]{ballesteros:00}%
  \BibitemOpen
  \bibfield  {author} {\bibinfo {author} {\bibfnamefont {H.~G.}\ \bibnamefont
  {Ballesteros}}, \bibinfo {author} {\bibfnamefont {A.}~\bibnamefont {Cruz}},
  \bibinfo {author} {\bibfnamefont {L.~A.}\ \bibnamefont {Fernandez}}, \bibinfo
  {author} {\bibfnamefont {V.}~\bibnamefont {Martin-Mayor}}, \bibinfo {author}
  {\bibfnamefont {J.}~\bibnamefont {Pech}}, \bibinfo {author} {\bibfnamefont
  {J.~J.}\ \bibnamefont {Ruiz-Lorenzo}}, \bibinfo {author} {\bibfnamefont
  {A.}~\bibnamefont {Tarancon}}, \bibinfo {author} {\bibfnamefont
  {P.}~\bibnamefont {Tellez}}, \bibinfo {author} {\bibfnamefont {C.~L.}\
  \bibnamefont {Ullod}}, \ and\ \bibinfo {author} {\bibfnamefont
  {C.}~\bibnamefont {Ungil}},\ }\href@noop {} {\bibfield  {journal} {\bibinfo
  {journal} {Phys. Rev. B}\ }\textbf {\bibinfo {volume} {62}},\ \bibinfo
  {pages} {14237} (\bibinfo {year} {2000})},\ \Eprint
  {http://arxiv.org/abs/arXiv:cond-mat/0006211} {arXiv:cond-mat/0006211}
  \BibitemShut {NoStop}%
\bibitem [{\citenamefont {Palassini}\ and\ \citenamefont
  {Caracciolo}(1999)}]{palassini:99}%
  \BibitemOpen
  \bibfield  {author} {\bibinfo {author} {\bibfnamefont {M.}~\bibnamefont
  {Palassini}}\ and\ \bibinfo {author} {\bibfnamefont {S.}~\bibnamefont
  {Caracciolo}},\ }\href@noop {} {\bibfield  {journal} {\bibinfo  {journal}
  {Phys. Rev. Lett.}\ }\textbf {\bibinfo {volume} {82}},\ \bibinfo {pages}
  {5128} (\bibinfo {year} {1999})},\ \Eprint
  {http://arxiv.org/abs/arXiv:cond-mat/9904246} {arXiv:cond-mat/9904246}
  \BibitemShut {NoStop}%
\bibitem [{\citenamefont {Bray}\ and\ \citenamefont {Moore}(2011)}]{bray:11}%
  \BibitemOpen
  \bibfield  {author} {\bibinfo {author} {\bibfnamefont {A.~J.}\ \bibnamefont
  {Bray}}\ and\ \bibinfo {author} {\bibfnamefont {M.~A.}\ \bibnamefont
  {Moore}},\ }\href@noop {} {\bibfield  {journal} {\bibinfo  {journal} {Phys.
  Rev. B}\ }\textbf {\bibinfo {volume} {83}},\ \bibinfo {pages} {224408}
  (\bibinfo {year} {2011})},\ \Eprint {http://arxiv.org/abs/arXiv:1102.1675}
  {arXiv:1102.1675} \BibitemShut {NoStop}%
\bibitem [{\citenamefont {Parisi}\ and\ \citenamefont
  {Temesv\'ari}(2011)}]{parisi:11}%
  \BibitemOpen
  \bibfield  {author} {\bibinfo {author} {\bibfnamefont {G.}~\bibnamefont
  {Parisi}}\ and\ \bibinfo {author} {\bibfnamefont {T.}~\bibnamefont
  {Temesv\'ari}},\ }\href@noop {} {\  (\bibinfo {year} {2011})},\ \Eprint
  {http://arxiv.org/abs/arXiv:1111.3313} {arXiv:1111.3313} \BibitemShut
  {NoStop}%
\bibitem [{\citenamefont {Bray}\ and\ \citenamefont {Roberts}(1980)}]{bray:80}%
  \BibitemOpen
  \bibfield  {author} {\bibinfo {author} {\bibfnamefont {A.~J.}\ \bibnamefont
  {Bray}}\ and\ \bibinfo {author} {\bibfnamefont {S.~A.}\ \bibnamefont
  {Roberts}},\ }\href@noop {} {\bibfield  {journal} {\bibinfo  {journal} {J.
  Phys. C: Solid St.Phys.}\ }\textbf {\bibinfo {volume} {13}},\ \bibinfo
  {pages} {5405} (\bibinfo {year} {1980})}\BibitemShut {NoStop}%
\bibitem [{\citenamefont {de~Almeida}\ and\ \citenamefont
  {Thouless}(1978)}]{dealmeida:78}%
  \BibitemOpen
  \bibfield  {author} {\bibinfo {author} {\bibfnamefont {J.~R.~L.}\
  \bibnamefont {de~Almeida}}\ and\ \bibinfo {author} {\bibfnamefont {D.~J.}\
  \bibnamefont {Thouless}},\ }\href@noop {} {\bibfield  {journal} {\bibinfo
  {journal} {J. Phys. A}\ }\textbf {\bibinfo {volume} {11}},\ \bibinfo {pages}
  {983} (\bibinfo {year} {1978})}\BibitemShut {NoStop}%
\bibitem [{Note1()}]{Note1}%
  \BibitemOpen
  \bibinfo {note} {A more careful analysis is needed in order to reach the same
  conclusion in the range $6<d<8$ \cite {fisher:85}.}\BibitemShut {Stop}%
\bibitem [{\citenamefont {Young}\ and\ \citenamefont
  {Katzgraber}(2004)}]{young:04}%
  \BibitemOpen
  \bibfield  {author} {\bibinfo {author} {\bibfnamefont {A.~P.}\ \bibnamefont
  {Young}}\ and\ \bibinfo {author} {\bibfnamefont {H.~G.}\ \bibnamefont
  {Katzgraber}},\ }\href@noop {} {\bibfield  {journal} {\bibinfo  {journal}
  {Phys. Rev. Lett.}\ }\textbf {\bibinfo {volume} {93}},\ \bibinfo {pages}
  {207203} (\bibinfo {year} {2004})},\ \Eprint
  {http://arxiv.org/abs/arXiv:cond-mat/0407031} {arXiv:cond-mat/0407031}
  \BibitemShut {NoStop}%
\bibitem [{\citenamefont {J\"org}\ \emph {et~al.}(2008)\citenamefont {J\"org},
  \citenamefont {Katzgraber},\ and\ \citenamefont {Krzakala}}]{jorg:08b}%
  \BibitemOpen
  \bibfield  {author} {\bibinfo {author} {\bibfnamefont {T.}~\bibnamefont
  {J\"org}}, \bibinfo {author} {\bibfnamefont {H.}~\bibnamefont {Katzgraber}},
  \ and\ \bibinfo {author} {\bibfnamefont {F.}~\bibnamefont {Krzakala}},\
  }\href@noop {} {\bibfield  {journal} {\bibinfo  {journal} {Phys. Rev. Lett.}\
  }\textbf {\bibinfo {volume} {100}},\ \bibinfo {pages} {197202} (\bibinfo
  {year} {2008})},\ \Eprint {http://arxiv.org/abs/arXiv:0712.2009}
  {arXiv:0712.2009} \BibitemShut {NoStop}%
\bibitem [{\citenamefont {J{\"o}nsson}\ \emph {et~al.}(2005)\citenamefont
  {J{\"o}nsson}, \citenamefont {Takayama}, \citenamefont {Aruga~Jatori},\ and\
  \citenamefont {Ito}}]{jonson:05}%
  \BibitemOpen
  \bibfield  {author} {\bibinfo {author} {\bibfnamefont {P.~E.}\ \bibnamefont
  {J{\"o}nsson}}, \bibinfo {author} {\bibfnamefont {H.}~\bibnamefont
  {Takayama}}, \bibinfo {author} {\bibfnamefont {H.}~\bibnamefont
  {Aruga~Jatori}}, \ and\ \bibinfo {author} {\bibfnamefont {A.}~\bibnamefont
  {Ito}},\ }\href@noop {} {\bibfield  {journal} {\bibinfo  {journal} {Phys.
  Rev. B}\ }\textbf {\bibinfo {volume} {71}},\ \bibinfo {pages} {180412(R)}
  (\bibinfo {year} {2005})},\ \Eprint
  {http://arxiv.org/abs/arXiv:cond-mat/0411291} {arXiv:cond-mat/0411291}
  \BibitemShut {NoStop}%
\bibitem [{\citenamefont {Petit}\ \emph {et~al.}(1999)\citenamefont {Petit},
  \citenamefont {Fruchter},\ and\ \citenamefont {Campbell}}]{petit:99}%
  \BibitemOpen
  \bibfield  {author} {\bibinfo {author} {\bibfnamefont {D.}~\bibnamefont
  {Petit}}, \bibinfo {author} {\bibfnamefont {L.}~\bibnamefont {Fruchter}}, \
  and\ \bibinfo {author} {\bibfnamefont {I.}~\bibnamefont {Campbell}},\
  }\href@noop {} {\bibfield  {journal} {\bibinfo  {journal} {Phys. Rev. Lett}\
  }\textbf {\bibinfo {volume} {83}},\ \bibinfo {pages} {5130} (\bibinfo {year}
  {1999})},\ \Eprint {http://arxiv.org/abs/arXiv:cond-mat/9910353}
  {arXiv:cond-mat/9910353} \BibitemShut {NoStop}%
\bibitem [{\citenamefont {Petit}\ \emph {et~al.}(2002)\citenamefont {Petit},
  \citenamefont {Fruchter},\ and\ \citenamefont {Campbell}}]{petit:02}%
  \BibitemOpen
  \bibfield  {author} {\bibinfo {author} {\bibfnamefont {D.}~\bibnamefont
  {Petit}}, \bibinfo {author} {\bibfnamefont {L.}~\bibnamefont {Fruchter}}, \
  and\ \bibinfo {author} {\bibfnamefont {I.}~\bibnamefont {Campbell}},\
  }\href@noop {} {\bibfield  {journal} {\bibinfo  {journal} {Phys. Rev. Lett}\
  }\textbf {\bibinfo {volume} {88}},\ \bibinfo {pages} {207206} (\bibinfo
  {year} {2002})},\ \Eprint {http://arxiv.org/abs/arXiv:cond-mat/011112}
  {arXiv:cond-mat/011112} \BibitemShut {NoStop}%
\bibitem [{\citenamefont {Tabata}\ \emph {et~al.}(2010)\citenamefont {Tabata},
  \citenamefont {Matsuda}, \citenamefont {Kanada}, \citenamefont {Yamazaki},
  \citenamefont {Waki}, \citenamefont {Nakamura}, \citenamefont {Sato},\ and\
  \citenamefont {Kindo}}]{tabata:10}%
  \BibitemOpen
  \bibfield  {author} {\bibinfo {author} {\bibfnamefont {Y.}~\bibnamefont
  {Tabata}}, \bibinfo {author} {\bibfnamefont {K.}~\bibnamefont {Matsuda}},
  \bibinfo {author} {\bibfnamefont {S.}~\bibnamefont {Kanada}}, \bibinfo
  {author} {\bibfnamefont {T.}~\bibnamefont {Yamazaki}}, \bibinfo {author}
  {\bibfnamefont {T.}~\bibnamefont {Waki}}, \bibinfo {author} {\bibfnamefont
  {H.}~\bibnamefont {Nakamura}}, \bibinfo {author} {\bibfnamefont
  {K.}~\bibnamefont {Sato}}, \ and\ \bibinfo {author} {\bibfnamefont
  {K.}~\bibnamefont {Kindo}},\ }\href@noop {} {\bibfield  {journal} {\bibinfo
  {journal} {Journal of Physical Society of Japan}\ }\textbf {\bibinfo {volume}
  {79}},\ \bibinfo {pages} {123704} (\bibinfo {year} {2010})},\ \Eprint
  {http://arxiv.org/abs/arXiv:1009.6115} {arXiv:1009.6115} \BibitemShut
  {NoStop}%
\bibitem [{\citenamefont {Moore}\ and\ \citenamefont
  {Drossel}(2002)}]{moore:02}%
  \BibitemOpen
  \bibfield  {author} {\bibinfo {author} {\bibfnamefont {M.}~\bibnamefont
  {Moore}}\ and\ \bibinfo {author} {\bibfnamefont {B.}~\bibnamefont
  {Drossel}},\ }\href@noop {} {\bibfield  {journal} {\bibinfo  {journal} {Phys.
  Rev. Lett.}\ }\textbf {\bibinfo {volume} {89}},\ \bibinfo {pages} {217202}
  (\bibinfo {year} {2002})},\ \Eprint
  {http://arxiv.org/abs/arXiv:cond-mat/0201107} {arXiv:cond-mat/0201107}
  \BibitemShut {NoStop}%
\bibitem [{\citenamefont {Leuzzi}\ \emph {et~al.}(2009)\citenamefont {Leuzzi},
  \citenamefont {Parisi}, \citenamefont {Ricci-Tersenghi},\ and\ \citenamefont
  {Ruiz-Lorenzo}}]{leuzzi:09}%
  \BibitemOpen
  \bibfield  {author} {\bibinfo {author} {\bibfnamefont {L.}~\bibnamefont
  {Leuzzi}}, \bibinfo {author} {\bibfnamefont {G.}~\bibnamefont {Parisi}},
  \bibinfo {author} {\bibfnamefont {F.}~\bibnamefont {Ricci-Tersenghi}}, \ and\
  \bibinfo {author} {\bibfnamefont {J.~J.}\ \bibnamefont {Ruiz-Lorenzo}},\
  }\href@noop {} {\bibfield  {journal} {\bibinfo  {journal} {Phys. Rev. Lett.}\
  }\textbf {\bibinfo {volume} {103}},\ \bibinfo {pages} {267201} (\bibinfo
  {year} {2009})},\ \Eprint {http://arxiv.org/abs/arXiv:0811.3435}
  {arXiv:0811.3435} \BibitemShut {NoStop}%
\bibitem [{\citenamefont {Fisher}\ and\ \citenamefont
  {Sompolinsky}(1985)}]{fisher:85}%
  \BibitemOpen
  \bibfield  {author} {\bibinfo {author} {\bibfnamefont {D.~S.}\ \bibnamefont
  {Fisher}}\ and\ \bibinfo {author} {\bibfnamefont {H.}~\bibnamefont
  {Sompolinsky}},\ }\href@noop {} {\bibfield  {journal} {\bibinfo  {journal}
  {Phys. Rev. Lett.}\ }\textbf {\bibinfo {volume} {54}},\ \bibinfo {pages}
  {1063} (\bibinfo {year} {1985})}\BibitemShut {NoStop}%
\bibitem [{\citenamefont {J{\"o}rg}\ and\ \citenamefont
  {Katzgraber}(2008)}]{jorg:08c}%
  \BibitemOpen
  \bibfield  {author} {\bibinfo {author} {\bibfnamefont {T.}~\bibnamefont
  {J{\"o}rg}}\ and\ \bibinfo {author} {\bibfnamefont {H.~G.}\ \bibnamefont
  {Katzgraber}},\ }\href@noop {} {\bibfield  {journal} {\bibinfo  {journal}
  {Phys. Rev. B}\ }\textbf {\bibinfo {volume} {77}},\ \bibinfo {pages} {214426}
  (\bibinfo {year} {2008})},\ \Eprint {http://arxiv.org/abs/arXiv:0803.3339}
  {arXiv:0803.3339} \BibitemShut {NoStop}%
\bibitem [{\citenamefont {Marinari}\ and\ \citenamefont
  {Zuliani}(1999)}]{marinari:99b}%
  \BibitemOpen
  \bibfield  {author} {\bibinfo {author} {\bibfnamefont {E.}~\bibnamefont
  {Marinari}}\ and\ \bibinfo {author} {\bibfnamefont {F.}~\bibnamefont
  {Zuliani}},\ }\href@noop {} {\bibfield  {journal} {\bibinfo  {journal} {J.
  Phys. A}\ }\textbf {\bibinfo {volume} {32}},\ \bibinfo {pages} {7447}
  (\bibinfo {year} {1999})},\ \Eprint
  {http://arxiv.org/abs/arXiv:cond-mat/9904303} {arXiv:cond-mat/9904303}
  \BibitemShut {NoStop}%
\bibitem [{\citenamefont {Amit}\ and\ \citenamefont
  {Martin-Mayor}(2005)}]{amit:05}%
  \BibitemOpen
  \bibfield  {author} {\bibinfo {author} {\bibfnamefont {D.~J.}\ \bibnamefont
  {Amit}}\ and\ \bibinfo {author} {\bibfnamefont {V.}~\bibnamefont
  {Martin-Mayor}},\ }\href@noop {} {\emph {\bibinfo {title} {Field Theory, the
  Renormalization Group and Critical Phenomena}}},\ \bibinfo {edition} {3rd}\
  ed.\ (\bibinfo  {publisher} {World Scientific},\ \bibinfo {address}
  {Singapore},\ \bibinfo {year} {2005})\BibitemShut {NoStop}%
\bibitem [{\citenamefont {Di~Francesco}\ \emph {et~al.}(1987)\citenamefont
  {Di~Francesco}, \citenamefont {Saleur},\ and\ \citenamefont
  {Zuber}}]{difrancesco:87}%
  \BibitemOpen
  \bibfield  {author} {\bibinfo {author} {\bibfnamefont {P.}~\bibnamefont
  {Di~Francesco}}, \bibinfo {author} {\bibfnamefont {H.}~\bibnamefont
  {Saleur}}, \ and\ \bibinfo {author} {\bibfnamefont {J.-B.}\ \bibnamefont
  {Zuber}},\ }\href@noop {} {\bibfield  {journal} {\bibinfo  {journal} {Nucl.
  Phys. B}\ }\textbf {\bibinfo {volume} {290}},\ \bibinfo {pages} {527}
  (\bibinfo {year} {1987})}\BibitemShut {NoStop}%
\bibitem [{\citenamefont {Di~Francesco}\ \emph {et~al.}(1988)\citenamefont
  {Di~Francesco}, \citenamefont {Saleur},\ and\ \citenamefont
  {Zuber}}]{difrancesco:88}%
  \BibitemOpen
  \bibfield  {author} {\bibinfo {author} {\bibfnamefont {P.}~\bibnamefont
  {Di~Francesco}}, \bibinfo {author} {\bibfnamefont {H.}~\bibnamefont
  {Saleur}}, \ and\ \bibinfo {author} {\bibfnamefont {J.-B.}\ \bibnamefont
  {Zuber}},\ }\href@noop {} {\bibfield  {journal} {\bibinfo  {journal}
  {Europhys. Lett.}\ }\textbf {\bibinfo {volume} {5}},\ \bibinfo {pages} {95}
  (\bibinfo {year} {1988})}\BibitemShut {NoStop}%
\bibitem [{\citenamefont {Ballesteros}\ \emph {et~al.}(1996)\citenamefont
  {Ballesteros}, \citenamefont {Fernandez}, \citenamefont {Martin-Mayor},\ and\
  \citenamefont {Mu{\~n}oz~Sudupe}}]{ballesteros:96}%
  \BibitemOpen
  \bibfield  {author} {\bibinfo {author} {\bibfnamefont {H.~G.}\ \bibnamefont
  {Ballesteros}}, \bibinfo {author} {\bibfnamefont {L.~A.}\ \bibnamefont
  {Fernandez}}, \bibinfo {author} {\bibfnamefont {V.}~\bibnamefont
  {Martin-Mayor}}, \ and\ \bibinfo {author} {\bibfnamefont {A.}~\bibnamefont
  {Mu{\~n}oz~Sudupe}},\ }\href@noop {} {\bibfield  {journal} {\bibinfo
  {journal} {Phys. Lett. B}\ }\textbf {\bibinfo {volume} {378}},\ \bibinfo
  {pages} {207} (\bibinfo {year} {1996})},\ \Eprint
  {http://arxiv.org/abs/arXiv:hep-lat/9511003} {arXiv:hep-lat/9511003}
  \BibitemShut {NoStop}%
\bibitem [{\citenamefont {Billoire}\ \emph {et~al.}(2011)\citenamefont
  {Billoire}, \citenamefont {Fernandez}, \citenamefont {Maiorano},
  \citenamefont {Marinari}, \citenamefont {Martin-Mayor},\ and\ \citenamefont
  {Yllanes}}]{billoire:11}%
  \BibitemOpen
  \bibfield  {author} {\bibinfo {author} {\bibfnamefont {A.}~\bibnamefont
  {Billoire}}, \bibinfo {author} {\bibfnamefont {L.~A.}\ \bibnamefont
  {Fernandez}}, \bibinfo {author} {\bibfnamefont {A.}~\bibnamefont {Maiorano}},
  \bibinfo {author} {\bibfnamefont {E.}~\bibnamefont {Marinari}}, \bibinfo
  {author} {\bibfnamefont {V.}~\bibnamefont {Martin-Mayor}}, \ and\ \bibinfo
  {author} {\bibfnamefont {D.}~\bibnamefont {Yllanes}},\ }\href@noop {}
  {\bibfield  {journal} {\bibinfo  {journal} {J. Stat. Mech.}\ (2011),\ \bibinfo
  {pages} {P10019}}},\ \Eprint
  {http://arxiv.org/abs/arXiv:1108.1336} {arXiv:1108.1336} \BibitemShut
  {NoStop}%
\bibitem [{\citenamefont {{\'A}lvarez~Ba{\~n}os}\ \emph
  {et~al.}(2010{\natexlab{a}})\citenamefont {{\'A}lvarez~Ba{\~n}os},
  \citenamefont {Cruz}, \citenamefont {Fernandez}, \citenamefont {Gil-Narvion},
  \citenamefont {Gordillo-Guerrero}, \citenamefont {Guidetti}, \citenamefont
  {Maiorano}, \citenamefont {Mantovani}, \citenamefont {Marinari},
  \citenamefont {Martin-Mayor}, \citenamefont {Monforte-Garcia}, \citenamefont
  {Mu{\~n}oz~Sudupe}, \citenamefont {Navarro}, \citenamefont {Parisi},
  \citenamefont {Perez-Gaviro}, \citenamefont {Ruiz-Lorenzo}, \citenamefont
  {Schifano}, \citenamefont {Seoane}, \citenamefont {Tarancon}, \citenamefont
  {Tripiccione},\ and\ \citenamefont {Yllanes}}]{janus:10b}%
  \BibitemOpen
  \bibfield  {author} {\bibinfo {author} {\bibfnamefont {R.}~\bibnamefont
  {{\'A}lvarez~Ba{\~n}os}}, \bibinfo {author} {\bibfnamefont {A.}~\bibnamefont
  {Cruz}}, \bibinfo {author} {\bibfnamefont {L.~A.}\ \bibnamefont {Fernandez}},
  \bibinfo {author} {\bibfnamefont {J.~M.}\ \bibnamefont {Gil-Narvion}},
  \bibinfo {author} {\bibfnamefont {A.}~\bibnamefont {Gordillo-Guerrero}},
  \bibinfo {author} {\bibfnamefont {M.}~\bibnamefont {Guidetti}}, \bibinfo
  {author} {\bibfnamefont {A.}~\bibnamefont {Maiorano}}, \bibinfo {author}
  {\bibfnamefont {F.}~\bibnamefont {Mantovani}}, \bibinfo {author}
  {\bibfnamefont {E.}~\bibnamefont {Marinari}}, \bibinfo {author}
  {\bibfnamefont {V.}~\bibnamefont {Martin-Mayor}}, \bibinfo {author}
  {\bibfnamefont {J.}~\bibnamefont {Monforte-Garcia}}, \bibinfo {author}
  {\bibfnamefont {A.}~\bibnamefont {Mu{\~n}oz~Sudupe}}, \bibinfo {author}
  {\bibfnamefont {D.}~\bibnamefont {Navarro}}, \bibinfo {author} {\bibfnamefont
  {G.}~\bibnamefont {Parisi}}, \bibinfo {author} {\bibfnamefont
  {S.}~\bibnamefont {Perez-Gaviro}}, \bibinfo {author} {\bibfnamefont
  {J.}~\bibnamefont {Ruiz-Lorenzo}}, \bibinfo {author} {\bibfnamefont {S.~F.}\
  \bibnamefont {Schifano}}, \bibinfo {author} {\bibfnamefont {B.}~\bibnamefont
  {Seoane}}, \bibinfo {author} {\bibfnamefont {A.}~\bibnamefont {Tarancon}},
  \bibinfo {author} {\bibfnamefont {R.}~\bibnamefont {Tripiccione}}, \ and\
  \bibinfo {author} {\bibfnamefont {D.}~\bibnamefont {Yllanes}} (\bibinfo
  {collaboration} {Janus Collaboration}),\ }\href@noop {} {\bibfield  {journal}
  {\bibinfo  {journal} {Phys. Rev. Lett.}\ }\textbf {\bibinfo {volume} {105}},\
  \bibinfo {pages} {177202} (\bibinfo {year} {2010}{\natexlab{a}})},\ \Eprint
  {http://arxiv.org/abs/arXiv:1003.2943} {arXiv:1003.2943} \BibitemShut
  {NoStop}%
\bibitem [{\citenamefont {Yllanes}(2011)}]{yllanes:11}%
  \BibitemOpen
  \bibfield  {author} {\bibinfo {author} {\bibfnamefont {D.}~\bibnamefont
  {Yllanes}},\ }\href@noop {} {\emph {\bibinfo {title} {Rugged Free-Energy
  Landscapes in Disordered Spin Systems}}}\ (\bibinfo  {publisher} {Ph.D.
  thesis, UCM},\ \bibinfo {year} {2011})\ \Eprint
  {http://arxiv.org/abs/arXiv:1111.0266} {arXiv:1111.0266} \BibitemShut
  {NoStop}%
\bibitem [{Note2()}]{Note2}%
  \BibitemOpen
  \bibinfo {note} {The data for $h=0.075$ presented very severe corrections,
  probably due to the proximity of the $h=0$ critical point. Therefore, we only
  use the data for $L\geq 12$ in order to estimate $T_\protect \text
  {c}(h=0.075)$.}\BibitemShut {Stop}%
\bibitem [{\citenamefont {Fisher}\ and\ \citenamefont
  {Huse}(1988)}]{fisher:88}%
  \BibitemOpen
  \bibfield  {author} {\bibinfo {author} {\bibfnamefont {D.~S.}\ \bibnamefont
  {Fisher}}\ and\ \bibinfo {author} {\bibfnamefont {D.~A.}\ \bibnamefont
  {Huse}},\ }\href@noop {} {\bibfield  {journal} {\bibinfo  {journal} {Phys.
  Rev. B}\ }\textbf {\bibinfo {volume} {38}},\ \bibinfo {pages} {373} (\bibinfo
  {year} {1988})}\BibitemShut {NoStop}%
\bibitem [{\citenamefont {Hartmann}(1999)}]{hartmann:99b}%
  \BibitemOpen
  \bibfield  {author} {\bibinfo {author} {\bibfnamefont {A.~K.}\ \bibnamefont
  {Hartmann}},\ }\href@noop {} {\bibfield  {journal} {\bibinfo  {journal}
  {Phys. Rev. E.}\ }\textbf {\bibinfo {volume} {60}},\ \bibinfo {pages} {5135}
  (\bibinfo {year} {1999})},\ \Eprint
  {http://arxiv.org/abs/arXiv:cond-mat/9904296} {arXiv:cond-mat/9904296}
  \BibitemShut {NoStop}%
\bibitem [{\citenamefont {Hukushima}(1999)}]{hukushima:99}%
  \BibitemOpen
  \bibfield  {author} {\bibinfo {author} {\bibfnamefont {K.}~\bibnamefont
  {Hukushima}},\ }\href@noop {} {\bibfield  {journal} {\bibinfo  {journal}
  {Phys. Rev. E}\ }\textbf {\bibinfo {volume} {60}},\ \bibinfo {pages} {3606}
  (\bibinfo {year} {1999})},\ \Eprint
  {http://arxiv.org/abs/arXiv:cond-mat/9903391} {arXiv:cond-mat/9903391}
  \BibitemShut {NoStop}%
\bibitem [{\citenamefont {Hukushima}\ and\ \citenamefont
  {Nemoto}(1996)}]{hukushima:96}%
  \BibitemOpen
  \bibfield  {author} {\bibinfo {author} {\bibfnamefont {K.}~\bibnamefont
  {Hukushima}}\ and\ \bibinfo {author} {\bibfnamefont {K.}~\bibnamefont
  {Nemoto}},\ }\href@noop {} {\bibfield  {journal} {\bibinfo  {journal} {J.
  Phys. Soc. Japan}\ }\textbf {\bibinfo {volume} {65}},\ \bibinfo {pages}
  {1604} (\bibinfo {year} {1996})},\ \Eprint
  {http://arxiv.org/abs/arXiv:cond-mat/9512035} {arXiv:cond-mat/9512035}
  \BibitemShut {NoStop}%
\bibitem [{\citenamefont {Marinari}(1998)}]{marinari:98b}%
  \BibitemOpen
  \bibfield  {author} {\bibinfo {author} {\bibfnamefont {E.}~\bibnamefont
  {Marinari}},\ }in\ \href@noop {} {\emph {\bibinfo {booktitle} {Advances in
  Computer Simulation}}},\ \bibinfo {editor} {edited by\ \bibinfo {editor}
  {\bibfnamefont {J.}~\bibnamefont {Kerst\'esz}}\ and\ \bibinfo {editor}
  {\bibfnamefont {I.}~\bibnamefont {Kondor}}}\ (\bibinfo  {publisher}
  {Springer-Berlag},\ \bibinfo {year} {1998})\BibitemShut {NoStop}%
\bibitem [{\citenamefont {Newman}\ and\ \citenamefont
  {Barkema}(1999)}]{newman:99}%
  \BibitemOpen
  \bibfield  {author} {\bibinfo {author} {\bibfnamefont {M.~E.~J.}\
  \bibnamefont {Newman}}\ and\ \bibinfo {author} {\bibfnamefont {G.~T.}\
  \bibnamefont {Barkema}},\ }\href@noop {} {\emph {\bibinfo {title} {{M}onte
  {C}arlo Methods in Statistical Physics}}}\ (\bibinfo  {publisher} {Clarendon
  Press},\ \bibinfo {address} {Oxford},\ \bibinfo {year} {1999})\BibitemShut
  {NoStop}%
\bibitem [{\citenamefont {{\'A}lvarez~Ba{\~n}os}\ \emph
  {et~al.}(2010{\natexlab{b}})\citenamefont {{\'A}lvarez~Ba{\~n}os},
  \citenamefont {Cruz}, \citenamefont {Fernandez}, \citenamefont {Gil-Narvion},
  \citenamefont {Gordillo-Guerrero}, \citenamefont {Guidetti}, \citenamefont
  {Maiorano}, \citenamefont {Mantovani}, \citenamefont {Marinari},
  \citenamefont {Martin-Mayor}, \citenamefont {Monforte-Garcia}, \citenamefont
  {Mu{\~n}oz~Sudupe}, \citenamefont {Navarro}, \citenamefont {Parisi},
  \citenamefont {Perez-Gaviro}, \citenamefont {Ruiz-Lorenzo}, \citenamefont
  {Schifano}, \citenamefont {Seoane}, \citenamefont {Tarancon}, \citenamefont
  {Tripiccione},\ and\ \citenamefont {Yllanes}}]{janus:10}%
  \BibitemOpen
  \bibfield  {author} {\bibinfo {author} {\bibfnamefont {R.}~\bibnamefont
  {{\'A}lvarez~Ba{\~n}os}}, \bibinfo {author} {\bibfnamefont {A.}~\bibnamefont
  {Cruz}}, \bibinfo {author} {\bibfnamefont {L.~A.}\ \bibnamefont {Fernandez}},
  \bibinfo {author} {\bibfnamefont {J.~M.}\ \bibnamefont {Gil-Narvion}},
  \bibinfo {author} {\bibfnamefont {A.}~\bibnamefont {Gordillo-Guerrero}},
  \bibinfo {author} {\bibfnamefont {M.}~\bibnamefont {Guidetti}}, \bibinfo
  {author} {\bibfnamefont {A.}~\bibnamefont {Maiorano}}, \bibinfo {author}
  {\bibfnamefont {F.}~\bibnamefont {Mantovani}}, \bibinfo {author}
  {\bibfnamefont {E.}~\bibnamefont {Marinari}}, \bibinfo {author}
  {\bibfnamefont {V.}~\bibnamefont {Martin-Mayor}}, \bibinfo {author}
  {\bibfnamefont {J.}~\bibnamefont {Monforte-Garcia}}, \bibinfo {author}
  {\bibfnamefont {A.}~\bibnamefont {Mu{\~n}oz~Sudupe}}, \bibinfo {author}
  {\bibfnamefont {D.}~\bibnamefont {Navarro}}, \bibinfo {author} {\bibfnamefont
  {G.}~\bibnamefont {Parisi}}, \bibinfo {author} {\bibfnamefont
  {S.}~\bibnamefont {Perez-Gaviro}}, \bibinfo {author} {\bibfnamefont
  {J.}~\bibnamefont {Ruiz-Lorenzo}}, \bibinfo {author} {\bibfnamefont {S.~F.}\
  \bibnamefont {Schifano}}, \bibinfo {author} {\bibfnamefont {B.}~\bibnamefont
  {Seoane}}, \bibinfo {author} {\bibfnamefont {A.}~\bibnamefont {Tarancon}},
  \bibinfo {author} {\bibfnamefont {R.}~\bibnamefont {Tripiccione}}, \ and\
  \bibinfo {author} {\bibfnamefont {D.}~\bibnamefont {Yllanes}} (\bibinfo
  {collaboration} {Janus Collaboration}),\ }\href@noop {} {\bibfield  {journal}
  {\bibinfo  {journal} {J. Stat. Mech.}\ (2010),\ \bibinfo {pages} {P06026}}
  },\ \Eprint
  {http://arxiv.org/abs/arXiv:1003.2569} {arXiv:1003.2569} \BibitemShut
  {NoStop}%
\bibitem [{\citenamefont {Cooper}\ \emph {et~al.}(1982)\citenamefont {Cooper},
  \citenamefont {Freedman},\ and\ \citenamefont {Preston}}]{cooper:82}%
  \BibitemOpen
  \bibfield  {author} {\bibinfo {author} {\bibfnamefont {F.}~\bibnamefont
  {Cooper}}, \bibinfo {author} {\bibfnamefont {B.}~\bibnamefont {Freedman}}, \
  and\ \bibinfo {author} {\bibfnamefont {D.}~\bibnamefont {Preston}},\
  }\href@noop {} {\bibfield  {journal} {\bibinfo  {journal} {Nucl. Phys. B}\
  }\textbf {\bibinfo {volume} {210}},\ \bibinfo {pages} {210} (\bibinfo {year}
  {1982})}\BibitemShut {NoStop}%
\bibitem [{\citenamefont {Zinn-Justin}(2005)}]{zinn-justin:05}%
  \BibitemOpen
  \bibfield  {author} {\bibinfo {author} {\bibfnamefont {J.}~\bibnamefont
  {Zinn-Justin}},\ }\href@noop {} {\emph {\bibinfo {title} {Quantum Field
  Theory and Critical Phenomena}}},\ \bibinfo {edition} {4th}\ ed.\ (\bibinfo
  {publisher} {Clarendon Press},\ \bibinfo {address} {Oxford},\ \bibinfo {year}
  {2005})\BibitemShut {NoStop}%
\bibitem [{\citenamefont {Bray}\ and\ \citenamefont {Moore}(1987)}]{bray:87}%
  \BibitemOpen
  \bibfield  {author} {\bibinfo {author} {\bibfnamefont {A.~J.}\ \bibnamefont
  {Bray}}\ and\ \bibinfo {author} {\bibfnamefont {M.~A.}\ \bibnamefont
  {Moore}},\ }in\ \href@noop {} {\emph {\bibinfo {booktitle} {Heidelberg
  Colloquium on Glassy Dynamics}}},\ \bibinfo {series and number} {\bibinfo
  {series} {Lecture Notes in Physics}\ No.\ \bibinfo {number} {275}},\ \bibinfo
  {editor} {edited by\ \bibinfo {editor} {\bibfnamefont {J.~L.}\ \bibnamefont
  {van Hemmen}}\ and\ \bibinfo {editor} {\bibfnamefont {I.}~\bibnamefont
  {Morgenstern}}}\ (\bibinfo  {publisher} {Springer},\ \bibinfo {address}
  {Berlin},\ \bibinfo {year} {1987})\BibitemShut {NoStop}%
\bibitem [{\citenamefont {de~Dominicis}\ \emph {et~al.}(1998)\citenamefont
  {de~Dominicis}, \citenamefont {Kondor},\ and\ \citenamefont
  {Temesv{\'a}ri}}]{dedominicis:98}%
  \BibitemOpen
  \bibfield  {author} {\bibinfo {author} {\bibfnamefont {C.}~\bibnamefont
  {de~Dominicis}}, \bibinfo {author} {\bibfnamefont {I.}~\bibnamefont
  {Kondor}}, \ and\ \bibinfo {author} {\bibfnamefont {T.}~\bibnamefont
  {Temesv{\'a}ri}},\ }in\ \href@noop {} {\emph {\bibinfo {booktitle} {{Spin
  {G}lasses and {R}andom {F}ields}}}},\ \bibinfo {editor} {edited by\ \bibinfo
  {editor} {\bibfnamefont {A.~P.}\ \bibnamefont {Young}}}\ (\bibinfo
  {publisher} {World Scientific},\ \bibinfo {address} {Singapore},\ \bibinfo
  {year} {1998})\BibitemShut {NoStop}%
\end{thebibliography}
\end{document}